\documentclass[journal]{IEEEtran}

\IEEEoverridecommandlockouts

\usepackage[utf8]{inputenc} 
\usepackage[T1]{fontenc}    
\usepackage{url}            
\usepackage{booktabs}       
\usepackage{amsfonts}       
\usepackage{nicefrac}       
\usepackage{microtype}      
\usepackage{xcolor}         
\usepackage[cmex10]{amsmath}

\usepackage{graphicx}   
\usepackage{caption} 
\usepackage{subcaption} 
\usepackage{makecell} 
\usepackage{framed}  
\usepackage{amsmath,amsfonts,amssymb,amsthm,epsfig,epstopdf,url,array}
\usepackage{cases}
\usepackage{bm}     
\usepackage{bbm}
\usepackage{cite}
\usepackage{algorithm} 
\usepackage{algorithmic}
\usepackage{color}
\usepackage{xparse}
\usepackage{mathtools}     
\usepackage{wasysym}
\usepackage[flushleft]{threeparttable}
\usepackage{kotex}
\newtheorem{theorem}{Theorem}
\newtheorem{lemma}{Lemma}

\newtheorem{assumption}{Assumption}

\def\BibTeX{{\rm B\kern-.05em{\sc i\kern-.025em b}\kern-.08em
    T\kern-.1667em\lower.7ex\hbox{E}\kern-.125emX}}

\usepackage{amsmath,amsfonts,bm}

\def\eqref#1{equation~\ref{#1}}

\def\1{\bm{1}}

\DeclareMathAlphabet{\mathsfit}{\encodingdefault}{\sfdefault}{m}{sl}
\SetMathAlphabet{\mathsfit}{bold}{\encodingdefault}{\sfdefault}{bx}{n}

\DeclareMathOperator*{\argmax}{arg\,max}

\begin{document} 
 
    \makeatletter
    \newcommand{\linebreakand}{%
      \end{@IEEEauthorhalign}
      \hfill\mbox{}\par
      \mbox{}\hfill\begin{@IEEEauthorhalign}
    }
    \makeatother

  \title{Cooperative Federated  Learning over Ground-to-Satellite  Integrated Networks: Joint Local Computation and Data Offloading}

\author{
Dong-Jun~Han,~\IEEEmembership{Member,~IEEE,}
Seyyedali Hosseinalipour,~\IEEEmembership{Member,~IEEE,}
David J. Love,~\IEEEmembership{Fellow,~IEEE,}   
Mung~Chiang,~\IEEEmembership{Fellow,~IEEE,}
Christopher G. Brinton,~\IEEEmembership{Senior Member,~IEEE}   
\thanks{This work was supported in part by  the Defense Advanced Research Projects Agency (DARPA)  under Grant D22AP00168, in part by the National Science Foundation (NSF) under Grant CNS-2212565, and in part by  the Office of
Naval Research (ONR)  under Grant N000142112472.

D.-J. Han, D. J. Love, M. Chiang and C. G. Brinton are with  the Elmore Family School of Electrical and Computer Engineering, 
Purdue University, West Lafayette, USA (e-mail: han762@purdue.edu; djlove@purdue.edu;
chiang@purdue.edu; cgb@purdue.edu).

S. Hosseinalipour is with the Department of Electrical Engineering,
University at Buffalo--SUNY, NY, USA (email: alipour@buffalo.edu).
}
}  
\maketitle

\begin{abstract}
While network coverage maps continue to expand, many devices located in remote areas remain unconnected to terrestrial communication infrastructures, preventing them from getting access to the associated data-driven services. In this paper, we propose a ground-to-satellite cooperative federated learning (FL) methodology to facilitate machine learning service management over remote regions. Our methodology orchestrates satellite constellations to provide the following key functions during FL: (i) processing data offloaded from ground devices, (ii) aggregating models within device clusters, and (iii) relaying models/data to other satellites via inter-satellite links (ISLs). 
Due to the limited coverage time of each satellite over a particular remote area, 
we facilitate satellite transmission of trained models and acquired data to neighboring satellites   via ISL, so that the incoming satellite can continue conducting FL for the region. We theoretically analyze the convergence behavior of our algorithm, and develop a training latency minimizer which optimizes over satellite-specific network resources, including the amount of data to be offloaded  from ground devices to satellites and   satellites' computation speeds.  
Through experiments on three   datasets, we show that our methodology can significantly speed up the convergence of FL compared with terrestrial-only and other satellite baseline approaches.
\end{abstract} 
\begin{IEEEkeywords}
Federated learning,  LEO satellites, ground-to-satellite integrated networks 
\end{IEEEkeywords}

 \bstctlcite{IEEEexample:BSTcontrol}

\section{Introduction}
In the era of big data, a plethora of valuable datasets are being collected by geo-distributed edge devices, such as smartphones, vehicles, and Internet-of-Things (IoT) sensors. Demand to leverage this data for intelligent services continues to grow. Motivated by this, federated learning (FL)  \cite{mcmahan2017communication, kairouz2021advances,
li2020federated_survey}  has been actively studied in recent years and has now become the de facto standard   for training a machine learning (ML) model  across   distributed nodes. By taking advantage of terrestrial communication infrastructures,  FL has  achieved a great success in both single server architectures \cite{mcmahan2017communication, wang2019adaptive, yang2019scheduling, amiri2020federated, chen2020joint, 
chen2020convergence} and hierarchical setups \cite{liu2020client, abad2020hierarchical, lim2021decentralized, lim2021dynamic} with multiple edge servers.  
  
\subsection{Motivation and Challenge}
\textbf{Motivation: FL across remote areas.}  Despite the wide deployment of mobile communication systems in current terrestrial networks, there still exist many  isolated regions on the Earth (e.g., rural regions,   maritime areas) that lack a well-developed communication infrastructure.  Since neither the servers nor the access points are available/reachable in these areas, it is currently difficult to conduct FL across   devices located in such regions. 
Although fully decentralized FL schemes have been   developed   
 \cite{wang2019matcha, roy2019braintorrent, lalitha2018fully,
koloskova2020unified}, which remove the reliance on edge servers, these methods   have limitations when  connections among devices are unstable (e.g.,  isolated regions with distant edge devices).

To facilitate collaborative training across these disconnected devices, we  take advantage of  Low Earth Orbit (LEO) satellites in  FL. The explosive growth in the number of LEO satellites enables the non-terrestrial network to cover most of the regions on the Earth, enabling satellites to work as a model aggregator during FL   in remote areas.     
 Especially when it is required to conduct FL across   multiple remote clusters that are geographically separated (as in Fig. \ref{fig:overview}), existing FL techniques, which mostly utilize terrestrial networks, face significant challenges as the devices in different regions are not able to aggregate their models  without the aid of non-terrestrial networks, e.g., satellites. Introducing satellites to FL also enables  low-powered ground devices  in isolated regions   to benefit from the computation capabilities of   satellites, due to the advancement of  on-board processing technology for satellites \cite{song2021energy, tang2021computation, cui2022latency,  li2021service, ding2021joint, zhang2023partial, tang2021deep}.

\textbf{Exemplary use cases.} Consider a scenario of training an ML model for natural disaster predictions (e.g., wildfire, hurricane, landslides). In order to reliably make decisions for these disasters, it is important to utilize data collected at IoT  sensors distributed across different  isolated areas, such as forests, oceans, and mountains.  
 Another key application could be autonomous vehicles, which necessitates collaboration among vehicles   in different  regions, including    rural areas  that lack a well-developed communication infrastructure.  Other examples include hospitals, smartphones, or wearable devices located in different rural regions  aiming to solve a classification task for diseases-prediction. These use cases motivate integrating   terrestrial networks  with the satellite network that has a wide coverage across  different remote regions  and sufficient computation powers  for   FL.

\textbf{Satellite-based computing/learning.} With the ever increasing development of space communication networks, LEO satellites are currently regarded as  a promising solution to achieve ubiquitous connection in 6G-and-beyond wireless.  The densely deployed LEO satellites enable   clients located in isolated regions to receive both high-quality communication and fast computation services.  Motivated by this, there has been an extensive work on satellite-assisted edge computing \cite{song2021energy, tang2021computation, cui2022latency,  li2021service, ding2021joint, zhang2023partial, tang2021deep}, where satellites process  tasks offloaded from low-powered  IoT devices. Although these works pave the way to integrate satellites in wireless networks, their focus is not on ML/FL.   More recently, researchers have become interested in conducting FL using data samples directly collected at satellites  \cite{so2022fedspace, 
 matthiesen2023federated, Razmi1, Razmi2, Razmi3, Elmahallawy1, zhai2023fedleo, Elmahallawy3} without consider.

\textbf{Key questions.} 
Despite   the above mentioned works,  utilization of satellites to assist   FL   across different remote terrestrial regions   has been less explored. In this work, we aim to explore this topic, which is faced with several key challenges and research questions. First, what should be the roles of the satellites in assisting FL across remote areas? How much data should be offloaded from the ground devices to the satellites in each region for  training? How should the satellite-side battery issue be handled during FL? How can we handle the limited coverage time of LEO satellites over each area? How can we guarantee convergence of a satellite-assisted FL algorithm? How should we optimize the network resources to   minimize the training time? To the best of our knowledge, this work is among the first attempts to address these research questions.

\subsection{Main Contributions}
In this paper, we  propose a ground-to-satellite cooperative FL methodology    that provides a novel direction for utilizing satellites in FL.   In order to facilitate efficient FL   across different remote areas or clusters, we use each satellite as (i) an edge computing unit for processing data offloaded from the ground users, (ii) a server for aggregating the models within each cluster, (iii) a relay for sharing the model/data with other satellites via inter-satellite links (ISLs).

Compared to the traditional  FL in terrestrial networks, one of the key challenges  that arises  in this ground-to-satellite integrated network  is that the satellites are continuously moving following their orbits, limiting the coverage time of satellites for each cluster.  To handle this issue,  we  propose  to let each satellite   transmit its trained model and data  offloaded from the clients to its neighboring satellite (that will cover the same region in a near future) via ISL, so that the incoming satellite can continue training the model for the devices located in that region.  We conduct convergence analysis of our FL algorithm, and optimize various network resources including the amount of data offloading, satellite-side computation power, and user bandwidth, to minimize the overall latency of ML model training. Another unique challenge of our setup is to deal with the satellite-side battery constraints. We consider solar-powered satellites  having solar panels where the batteries could be charged via the sun. Hence, more data can be offloaded to the satellites when clusters are facing the sun, and those satellites can process data with higher computation powers with less battery issues. These aspects are integrated in our formulated network optimization problem.

Our main contributions can be summarized as follows:  
 \begin{itemize}
\item We propose a satellite-assisted   FL methodology    that enables ground users in different remote areas to collaboratively train an ML model, based on hybrid client-side local computation, ground-to-satellite data offloading, satellite-side model training,
satellite-assisted intra-cluster aggregation,  and satellite-to-satellite model/data transmissions.
\item Based on the objective function  defined with satellite-side and client-side local losses in each cluster, we theoretically  analyze the convergence behavior of our FL algorithm  with respect to the amount of data processed at individual satellites and ground devices, and guarantee its convergence to a stationary point of general non-convex loss functions.
\item We formulate a network optimization problem to minimize latency, and optimize  satellite-specific key design elements including the amount of data to be offloaded from each ground user to each satellite and satellites' computation powers, while considering the battery constraints of solar-powered satellites. 
\item In simulations, we adopt three  benchmark datasets for FL  and  show that our methodology can significantly speed up the convergence of FL compared with  baselines  by strategically taking advantage of satellites' resources.  We also provide key insights into the effects of data offloading depending on whether the cluster is facing the sun or not.  
\end{itemize}
To the best of our knowledge, this is one of the earliest works to provide an analysis on \textit{FL over ground-to-satellite  integrated  networks}  with theoretical guarantee and optimized design elements. Our solution provides new guidelines on how to utilize satellites in FL, with several unique characteristics including the impact of  battery charging of solar-powered   satellites, and  satellite-to-satellite model/data transmissions. 
 
 \subsection{Related Works}\label{sec:related}

\textbf{FL over terrestrial  networks.} 
Most existing works on FL have been developed in a terrestrial network  with a single cloud server \cite{mcmahan2017communication, wang2019adaptive, yang2019scheduling, amiri2020federated, chen2020joint, 
chen2020convergence, han2023federated} or with multiple edge servers \cite{liu2020client, abad2020hierarchical, lim2021decentralized, lim2021dynamic, han2021fedmes}.  In these works, the role  of the server is to aggregate the models sent from the clients. Research has been also conducted in a fully decentralized setup \cite{wang2019matcha, roy2019braintorrent, lalitha2018fully,
koloskova2020unified} where the models are aggregated at individual clients, without  requiring any servers. However, when clients in different  remote regions   need to collaboratively train an ML model without relying on terrestrial communications,  these FL techniques face  great challenges.


\textbf{FL over satellites/UAVs.}  Motivated by the recent proliferation of LEO satellites and the advancement of their computation capabilities, there has  been  a line of works focusing on FL over satellites \cite{matthiesen2023federated, so2022fedspace, Razmi1, Razmi2, Razmi3, Elmahallawy1, zhai2023fedleo, Elmahallawy3}, where data samples are directly collected at the satellites. The satellites can be viewed as clients, and perform local updates using their local datasets.   
The main focus of these works is to investigate where/how to aggregate the trained models after the satellite-side local updates. 
The authors of  \cite{so2022fedspace, Razmi1, Razmi2, Razmi3, Elmahallawy3}  specifically study  model aggregation strategies by utilizing  the   ground station as a server,  where \cite{Razmi1,  Elmahallawy3} specifically adopt  ISL communications during the aggregation process.  
FL has been  also studied in unmanned aerial vehicle (UAV) networks, where the UAVs  collect data and act as clients \cite{wang2020learning,zhang2020federated,zeng2020federated}. Compared to these works where either the LEO satellites or the UAVs  are viewed as clients, we focus on a different scenario where clients are located in   isolated regions on the ground.    Our problem setup necessitates satellites to have additional functions to process data offloaded from the ground users, to aggregate the models within the cluster, and to relay model/data to another satellite via ISL  communications to continue model training for the cluster.

\textbf{Satellite-assisted  computing/learning for ground users.}
Another line of works focus on satellite-based edge computing \cite{song2021energy, tang2021computation, cui2022latency,  li2021service, ding2021joint, zhang2023partial, tang2021deep} to develop computation offloading strategies to the satellites. However, their focus is not on ML/FL. 
The key difference between these literatures and our setup is that, we require    clients to  collaboratively   train a shared ML model across different regions, while the clients in prior works just need to solve their  own computational tasks independently. The collaborative ML task in our setup requires satellites to aggregate and relay models to theoretically guarantee the convergence of FL, making the problem totally different.

 Only a few prior works \cite{rodrigues2023hybrid, 
 chen2022satellite, fang2022olive, wang2022federated} have focused on satellite-assisted distributed ML     as in our setup, where the ground users  aim at constructing an ML model assisted by the satellites. This problem setup is fundamentally different from the settings in \cite{matthiesen2023federated, so2022fedspace, Razmi1, Razmi2, Razmi3, Elmahallawy1,  zhai2023fedleo, Elmahallawy3} where each satellite directly collects its local data, presenting new research questions on what the roles of the satellites should be.   
The recent magazine paper \cite{chen2022satellite}  discusses possible scenarios of utilizing satellites as edge computing servers or relays during FL. In \cite{rodrigues2023hybrid}, the authors specifically focus on training the deep Q network (DQN) model  for
optimizing the action selection policy in the maze problem. In \cite{fang2022olive, wang2022federated}, the satellite is also adopted  as a server to aggregate the models trained at the ground devices in wireless settings.  In \cite{wang2022uav}, the authors consider data offloading in UAV-assisted  FL. 
 Compared to prior works, our contribution is to design a  satellite-assisted FL strategy     with several unique characteristics including client-to-satellite data offloading, hybrid satellite/client model training,  battery charging of solar-powered LEO satellites, and   model/data transmissions between incoming and departing satellites.

\subsection{Organization}
The rest of this paper is organized as follows.  Section \ref{sec:system} illustrates the system model and overview of  approach, while Section \ref{sec:main} presents our  satellite-assisted FL algorithm. Theoretical convergence analysis and  network optimization results are provided in Section \ref{sec:convergence} and Section \ref{sec:optimization}, respectively.  We present experimental results in Section \ref{sec:experiments}, and draw conclusions in Section \ref{sec:conclusion}.

\section{System Model and Overview of Approach} \label{sec:system}

Consider a set  $\mathcal{J}$ of $J$ different clusters  located in different  remote areas that are geographically separated.   We also  consider a set $\mathcal{K}$ of $K$ users or clients distributed across these $J$ clusters, where each user belongs  to a single cluster.  Fig. \ref{fig:overview} shows an example with $J=3$ clusters and $K=9$ clients. Let $G_j$ be the set of client indices located in cluster $j\in \mathcal{J}$. The users in these clusters are  not supported by terrestrial communication infrastructures. In order to   facilitate FL across clients located in these  remote regions, we consider LEO satellites moving around the Earth following their own orbits. Individual clusters can be supported by the satellites from different orbits, or possibly the same orbit depending on their locations. Each client $k$ has its own local dataset $D_k= D_k^c \cup D_k^s$ with $D_k^c \cap D_k^s=\emptyset$,  where $D_k^c$ is the set of privacy-sensitive samples of client $k$ that should be necessarily kept in each \textit{client}, while $D_k^s$ is the set of non-sensitive samples   that can be possibly offloaded to the   \textit{satellite}.   The sensitive and non-sensitive data samples can be distinguished at each client based on the (i) locations of collected data, (ii) class information of data, or (iii) some other predetermined information depending on the specific task.

\textbf{Solar-powered satellites.} In each time slot, the overall  $J$ cluster sets $\{G_j\}_{j=1}^J$ in the system  can be divided into two different categories depending on their locations.    First are the  clusters  in which the satellites covering the region can charge their batteries via the sun. This becomes possible when  cluster $j$ is facing the sun and the corresponding satellites' solar panels have no orientation issues for charging. We define   $\mathcal{J} ^{\text{sun}}$ as the set of    the corresponding cluster indices: The batteries of   LEO satellites covering the clusters in $j\in\mathcal{J}^{\text{sun}}$ will have  less battery issues at the satellites due to the solar power. On the other hand, the batteries of satellites covering the remaining clusters (i.e., $j\in\mathcal{J}
\setminus\mathcal{J}^{\text{sun}}$) are not able to be charged, which will result in relatively strict energy constraints.


\textbf{Goal and overview of approach.} Under this system model, the goal of the ground users is to construct a shared global model $\mathbf{w}^*$ that well reflects all data samples in $J$ clusters in the system. Fig. \ref{fig:overview} describes the high-level overview of our approach. As a preprocessing step, the resource-limited clients offload specific portions of their local datasets to the corresponding satellite as in Fig. \ref{fig:overview_a}. Then in Fig. \ref{fig:overview_b}, each client performs local updates using its remaining data samples, while each satellite performs local updates using   data offloaded from the clients. Here, due to the limited coverage time of each satellite over each cluster, we let each satellite  transmits  the trained model and the collected dataset to the neighboring satellite that will cover the same cluster  in the near future. As a result, the   incoming satellite can continue training the model for the specific cluster. This procedure is repeated until all  data samples at the satellite are processed. Then in Fig. \ref{fig:overview_c}, intra-cluster model aggregation is performed to construct cluster-specific models.  Finally, in Fig. \ref{fig:overview_d}, a global model is constructed  by aggregating the models of all clusters,  which is sent back to all clients at the beginning of next global round. 


%
%

\begin{figure*}[t]
  \centering
 \begin{subfigure}[b]{0.46\textwidth}
         \centering
         \includegraphics[width=\textwidth]{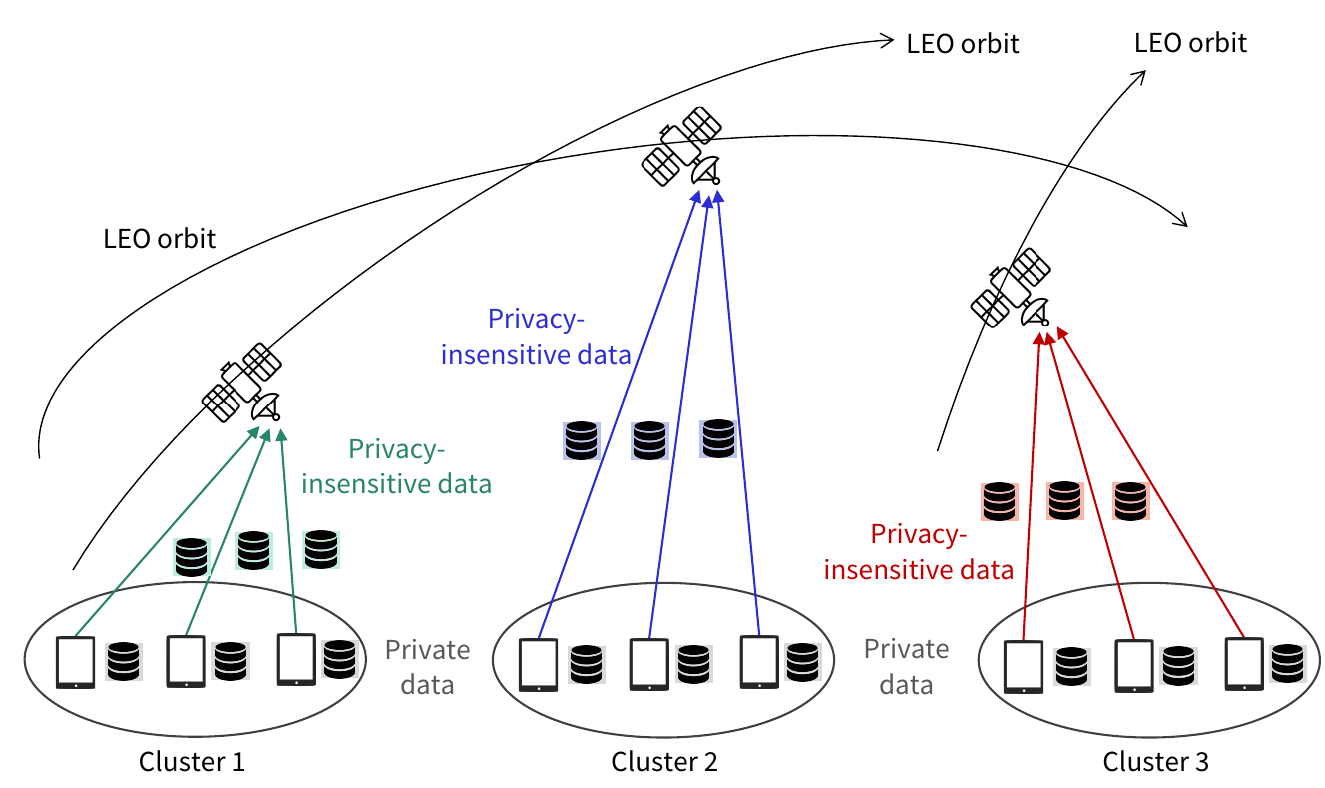} 
     \caption{Preprocessing: Data offload}\label{fig:overview_a}
 \end{subfigure}  \hspace{5mm} 
   \begin{subfigure}[b]{0.46\textwidth}
         \centering
         \includegraphics[width=\textwidth]{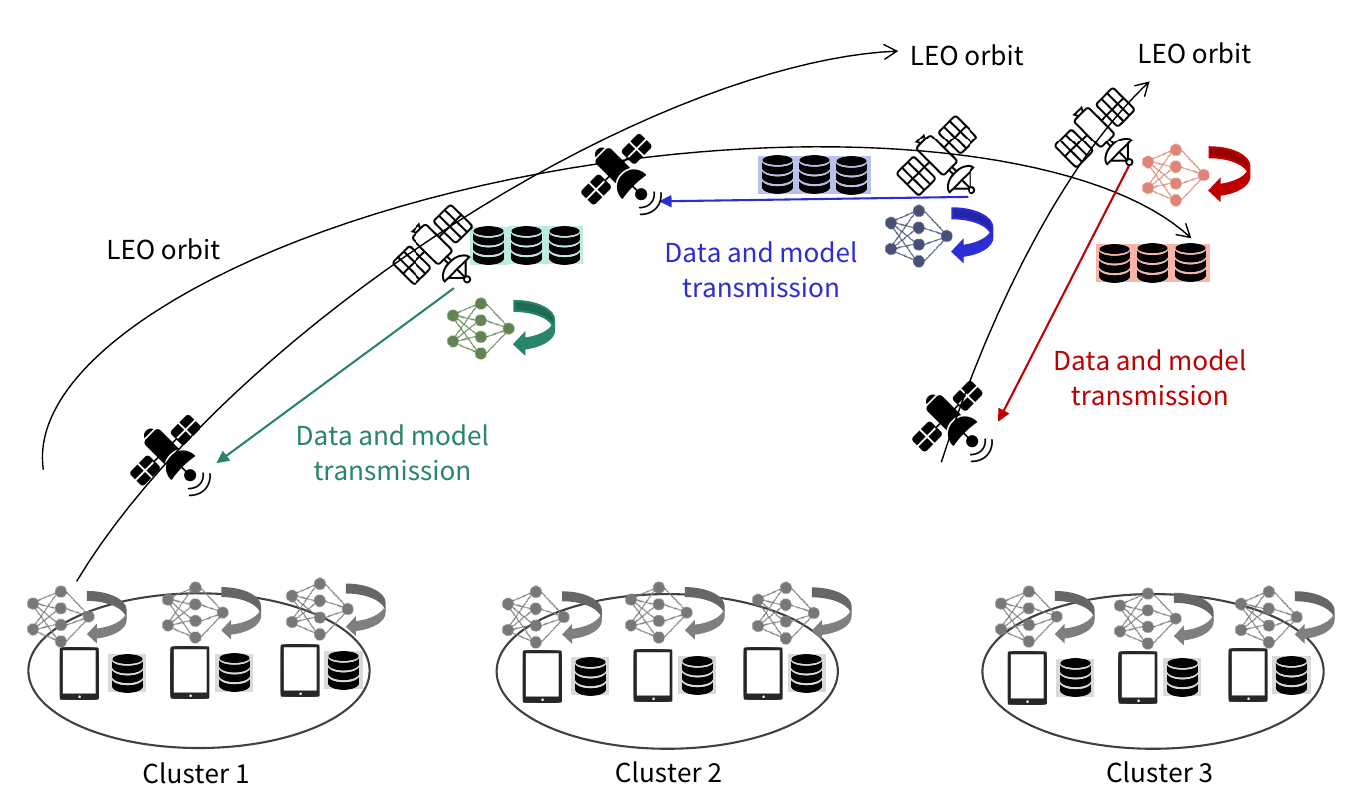} 
     \caption{Local update and data/model transmission via ISL}\label{fig:overview_b}
 \end{subfigure}
  \begin{subfigure}[b]{0.46\textwidth}
         \centering
         \includegraphics[width=\textwidth]{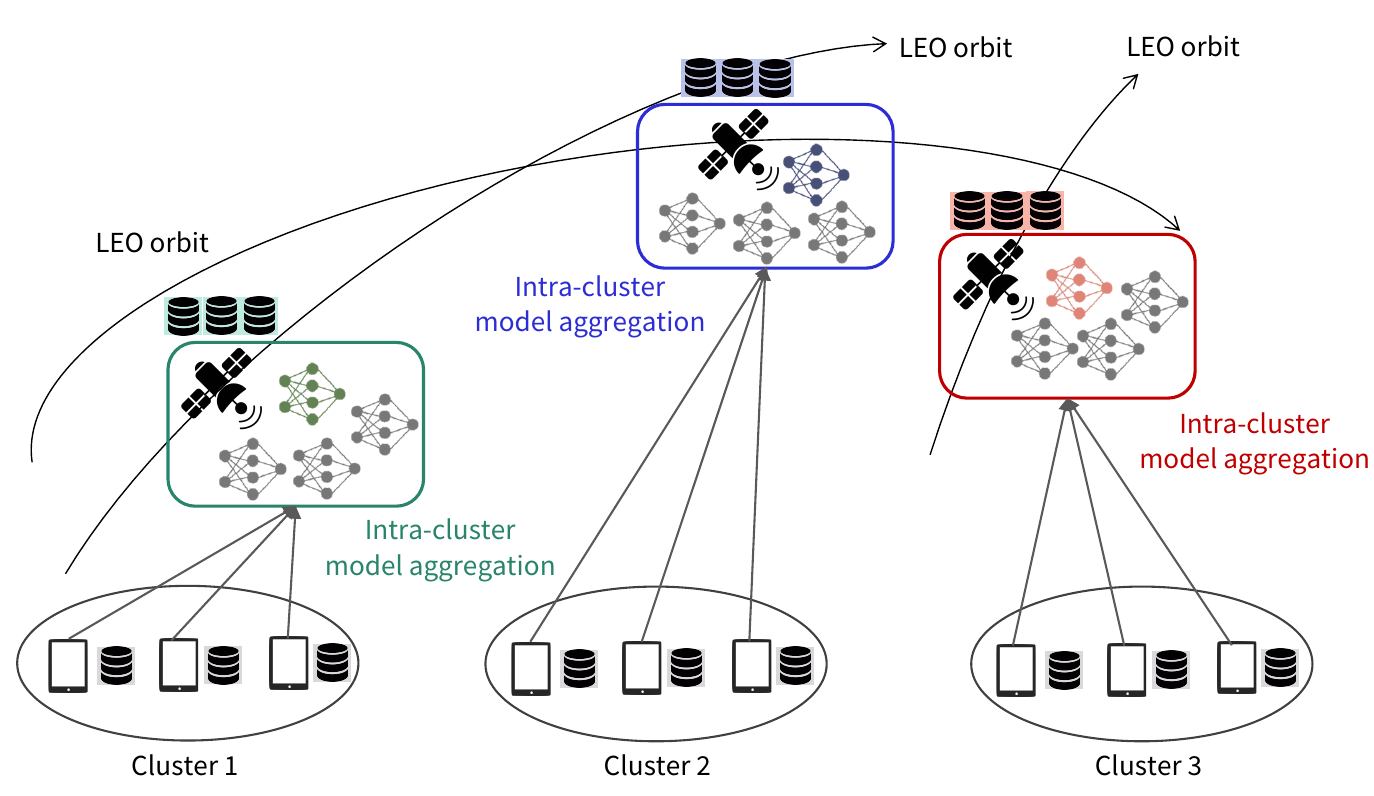} 
     \caption{Intra-cluster model aggregation}\label{fig:overview_c}
 \end{subfigure} \hspace{3mm} 
   \begin{subfigure}[b]{0.46\textwidth}
         \centering
         \includegraphics[width=\textwidth]{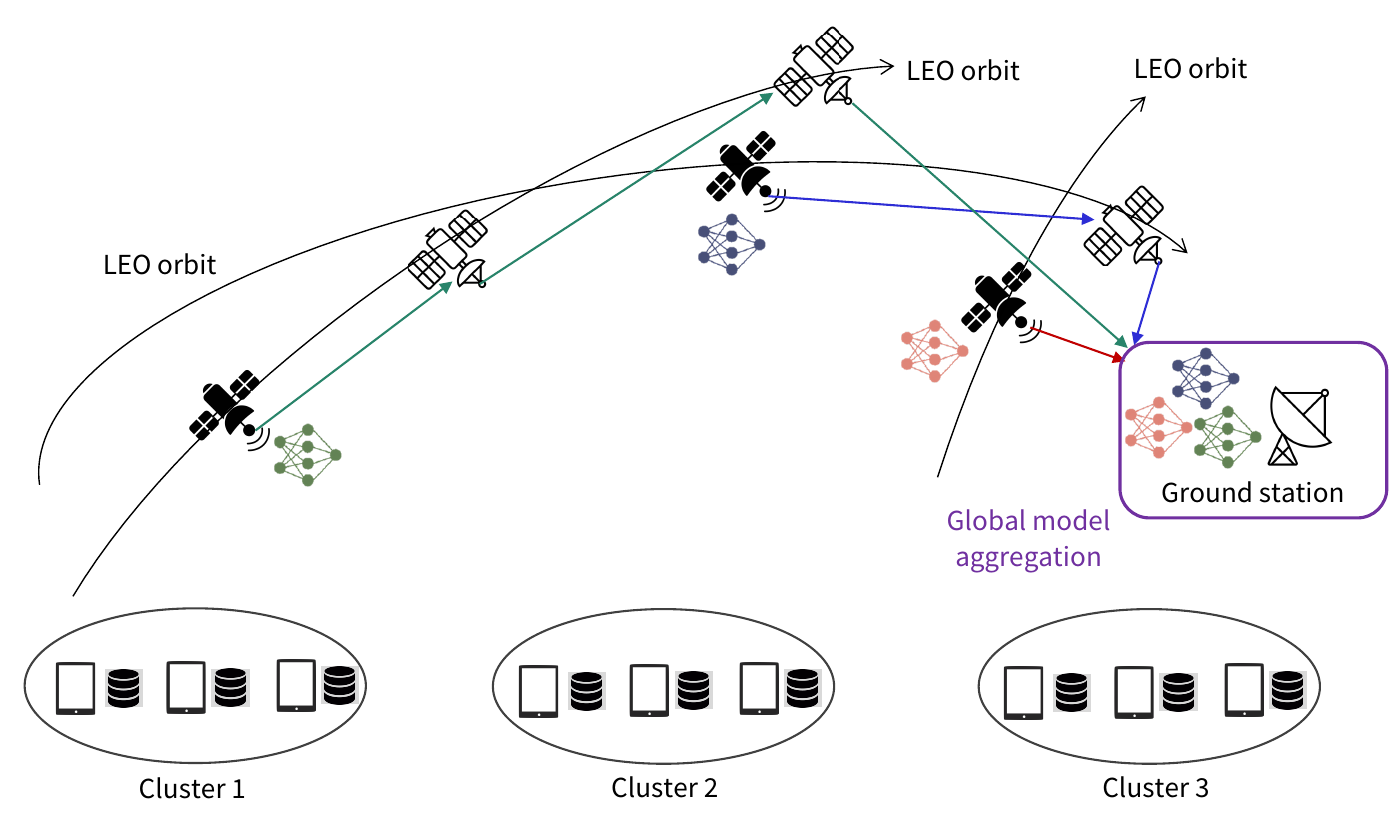} 
     \caption{Global model aggregation}\label{fig:overview_d}
 \end{subfigure}
     \caption{Overview of proposed satellite-assisted FL algorithm. The preprocessing step in Fig. \ref{fig:overview_a} is conducted only once before training begins, while the other steps in Figs. \ref{fig:overview_b}, \ref{fig:overview_c}, \ref{fig:overview_d} are repeated for $R$ global  FL rounds.}
    \label{fig:overview}
    \vspace{-2mm}
  \end{figure*}

 \section{Proposed Algorithm} \label{sec:main}
The step in Fig. \ref{fig:overview_a} is a preprocessing step that occurs only once before training begins, while the other steps in Figs. \ref{fig:overview_b}, \ref{fig:overview_c},  \ref{fig:overview_d} are   repeated for $R$ global rounds, which we   index $r=0, 1, \dots, R-1$.  
  In each round $r$, the global model $\mathbf{w}^r$ is constructed by synchronously aggregating the satellite models at the ground station as in \cite{Razmi1, Elmahallawy3}.    We propose all LEO satellites to play the following key  roles simultaneously during training: Each satellite  work as an (i) \textit{edge computing unit} for updating the model using data samples offloaded from the clients, as  a (ii) \textit{server} for intra-cluster model aggregation, and as a (iii) \textit{relay} for sharing the updated model and data to other satellites via ISL communications.
Based on these unique characteristics of our framework,   in the following, we will describe our methodology    with details.

\subsection{Preprocessing step:   Data Offload  to Satellites} 
 \label{subsec:preproc}
Consider  a specific cluster set $G_j$. Before training begins,  as a preprocessing step,  each client $k\in G_j$ offloads a subset of non-sensitive dataset $\hat{D}_k^s \subset D_k^s$ to the corresponding   satellite that is currently covering cluster $j$, as shown in Fig. \ref{fig:overview_a}.  Here, we   introduce our key design parameter 
\begin{equation}\label{eq:alpha_def}
\alpha_k = \frac{|\hat{D}_k^s|}{|D_k|},
\end{equation}
which is   the portion of data samples offloaded from client $k$ to the satellite. Here,  for an arbitrary dataset $D$, we use $|D|$ to denote the number of data samples in  $D$. We  have $0\leq\alpha_k\leq \alpha_k^{\text{max}}$, where $ \alpha_k^{\text{max}} =|D_k^s|/|D_k|$ is the portion of non-sensitive samples in client $k$ which may depend on the specific ML application (e.g., IoT sensors  for disaster predictions may have large $\alpha_k^{\text{max}}$ while personal vehicles may have smaller $\alpha_k^{\text{max}}$). 


\textbf{Client-side objective function.} After the data offloading process, the $k$-th client's objective  function is defined with the remaining dataset $\hat{D}_k^c = D_k\setminus\hat{D}_k^s$  as follows:
\vspace{-0.5mm}
\begin{align}\label{eq:client_objective}
\ell_{C,k}(\mathbf{w})&=\frac{1}{|\hat{D}_k^c |}\sum_{x\in \hat{D}_k^c }\ell(x;\mathbf{w})\\  &= \frac{1}{(1-\alpha_k)|D_k |}\sum_{x\in \hat{D}_k^c }\ell(x;\mathbf{w}),
\end{align}
where $\ell(x; \mathbf{w})$ is the loss computed with data sample $x$ and model $\mathbf{w}$.  In $\ell_{C,k}(\mathbf{w})$, $C$ stands for the client and $k$ is the index of the client.

\textbf{Satellite-side objective function.}  On the other hand, the  satellite-side objective function for cluster $j$ is defined with the  dataset $\cup_{k\in G_j} \hat{D}_k^s$ offloaded from the clients in $G_j$ as
\vspace{-0.5mm}
\begin{align}\label{eq:satellite_objective}
\ell_{S,j}(\mathbf{w})&=\frac{1}{|\cup_{k\in G_j} \hat{D}_k^s|}\sum_{x\in \cup_{k\in G_j} \hat{D}_k^s}\ell(x;\mathbf{w})\\ &=\frac{1}{\sum_{k\in G_j}\alpha_k |D_k|}\sum_{x\in \cup_{k\in G_j} \hat{D}_k^s}\ell(x;\mathbf{w}),
\end{align}
where $S$ stands for satellite and $j$ is the index of the cluster.


 \subsection{Step 1: Client-Side Local Training}\label{sec:step2}
In the beginning of each round $r$, all clients and satellites have model $\mathbf{w}^r$.  As in Fig. \ref{fig:overview_b}, 
each client $k$ locally trains the model using $\hat{D}_k^c = D_k\setminus\hat{D}_k^s$, which is the remaining dataset  after the data offloading step in Section \ref{subsec:preproc}.  Here, we introduce another  variable $\gamma_k^r$, which is the portion of data samples that client $k$ processes during the model update at  a specific round $r$. By defining $\lambda_{C,k}^r$ as the number of samples  being processed at client $k$ at a specific round $r$, we can write 
\begin{equation}\label{eq:gamma}
\gamma_k^r = \frac{\lambda_{C,k}^r}{|D_k|}.
\end{equation}
Here, since $0<\lambda_{C,k}^r\leq |\hat{D}_k^c|  = (1-\alpha_k)|D_k|$ holds, we  have $0 < \gamma_k^r \leq 1-\alpha_k$. One possible solution is to let each client $k$  process all the remaining data samples by setting $\gamma_k^r  = 1-\alpha_k$. In specific time-sensitive applications, one can also set a smaller $\gamma_k^r$ to further reduce the training time at low-powered IoT devices/sensors in remote regions. Taking a small $\gamma_k^r$ will reduce the training time at each device, with the cost of degraded learning performance, as we will see later in the convergence analysis in Section \ref{sec:convergence}. 
For each client $k$, the local   update process becomes
 \begin{equation}\label{eq:clientlocaljupdate}
\mathbf{w}^{r+1}_{C, k} = \mathbf{w}^{r} - \eta_r  \tilde{\nabla} \ell_{C, k}(\mathbf{w}^r),
\end{equation}
where $\eta_r$ is the learning rate at round $r$ and  $\tilde{\nabla}\ell_{C, k}(\mathbf{w})$ 
 is the gradient computed with a specific mini-batch of size  $\lambda_{C,k}^r$ during the mini-batch stochastic gradient descent (SGD) update.

The computation time $\tau_{C,k}^{\text{local}}$ (in seconds) to finish the local update at client $k$ is   written as
\begin{equation}
\tau_{C,k}^{\text{local}} = \frac{m_{C,k} \gamma_k^r |D_k|}{f_{C,k}},
\end{equation}
where  $f_{C,k}$ (in $\text{cycles}/\text{sec}$ or Hz)  is the CPU frequency at client $k$ and $m_{C,k}$ (in $\text{cycles}/\text{sample}$) is the number of CPU cycles for the $k$-th client to process one data sample.  Here, $m_{C,k}$ is  a fixed value that can be obtained offline at each client $k$  as assumed in  \cite{yang2020energy, 
dinh2020federated}.  We can also write the corresponding energy consumption  $E_{C,k}^{\text{local}}$ (in Joules) as 
\begin{equation}
E_{C,k}^{\text{local}} = \kappa  m_{C,k}  \gamma_k^r  |D_k| f_{C,k}^2 ,
\end{equation}
where $\kappa$ (in $\text{Joules}\cdot \text{seconds}^2/\text{cycles}^3$) is the energy consumption coefficient.

\subsection{Step 2: Repeated Satellite-Side Computation and ISL  Communication (working in parallel with Step 1)}

While the clients are performing local updates in  Step 1, the satellite covering cluster $j$ also performs model update based on the  dataset $\cup_{k\in G_j} \hat{D}_k^s$  offloaded from the clients in  the cluster set $G_j$.  We let the satellite that is covering cluster $j$   perform local updates until it processes all the  data samples in $\cup_{k\in G_j} \hat{D}_k^s$.  
However, since each satellite is continuously moving following its own orbit, the coverage time of the satellite over each cluster limited. As a result,  the satellite may not finish the computation before leaving   cluster $j$. We let this coverage time  of the satellite for each cluster to be $T$\footnote{In practice, the coverage time provided by each successive satellite   varies over time according to the satellite constellation model. In such cases, $T$ can be obtained by taking the average of coverage times of future satellites and used for optimization. We show later  that this approach   works well in practical constellation models. Exploring a precise  optimization algorithm considering varying coverage times is an interesting direction for future research.}. Whenever a specific satellite leaves cluster $j$ after   $T$ seconds, a new satellite in the same or adjacent orbit will start covering cluster $j$. Motivated by this  characteristic of the satellite network, we propose to let the satellite   transmit the collected dataset $\cup_{k\in G_j} \hat{D}_k^s$ and the trained model to the new satellite before leaving cluster $j$, so that the incoming satellite can continue the training process for cluster $j$. When the last satellite for cluster $j$ finishes  computation  for the global round, it   transmits the data/model to the new satellite for the next global round. Fig. \ref{fig:overview_b} describes this process in a high-level. Specifically, our idea is to repeat the below two steps multiple times until all the samples in $\cup_{k\in G_j} \hat{D}_k^s$ are processed.

\textbf{Step 2-1: Transmissions of data and model via ISL.} Before a satellite joins   cluster $j$, it receives the model and data from the previous  satellite that is leaving cluster $j$. The communication time for this process can be written as
\begin{equation}\label{eq:tau_trans}
\tau_{S,j}^{\text{trans}} = \frac{S(\mathbf{w})+ q\sum_{k\in G_j}\alpha_k |D_k| }{Q_j^{\text{ISL}}},
\end{equation}
 where $S(\mathbf{w})$ denotes the size of the model (in bits), $q$  is the size of each data sample (in bits),  $Q_j^{\text{ISL}}$ is the transmission rate for ISL between two adjacent satellites in cluster $j$.  Here,  $S(\mathbf{w})$ can be calculated based on the number of model parameters in $\mathbf{w}$ and the type of quantization  applied to individual   parameters  (e.g., $S(\mathbf{w})= 32\times |\mathbf{w}|$ upon using 32 bit quantization). Following \cite{leyva2021inter, Razmi1}, we can also write $Q_j^{\text{ISL}}=\mathcal{B}_j\log_2(1+SNR_j)$, where  $\mathcal{B}_j$ is the channel bandwidth and $SNR_j=\frac{p_{S,j}G_j^{\text{rx}}G_j^{\text{tx}}}{ZN_0}$ is the signal to noise ratio between two satellites. Here, $G_j^{\text{rx}}$, $G_j^{\text{tx}}$, $Z$, $N_0$, $p_{S,j}$ are the  Rx, Tx gains of the antenna, free space path loss,  noise power density, and  transmit power at the satellite, respectively.  The corresponding energy consumption for  this ISL communication procedure (at the satellite that is leaving cluster $j$)  can be written as follows:
\begin{equation}\label{eq:Energy_trans}
E_{S,j}^{\text{trans}} = p_{S,j} \cdot \tau_{S,j}^{\text{trans}}.
\end{equation}

\textbf{Step 2-2: Satellite-side local update.} After receiving the model and data, the new satellite covering cluster $j$ performs local updates using dataset $\cup_{k\in G_j} \hat{D}_k^s$   according to
\begin{equation}
\mathbf{w}^{r+1}_{S, j} = \mathbf{w}^{r} - \eta_r  \tilde{\nabla} \ell_{S, j}(\mathbf{w}^r),
\end{equation}
where $\tilde{\nabla}\ell_{S, j}(\mathbf{w})$ is the gradient of the mini-batch defined with (\ref{eq:satellite_objective}).  The model update is conducted until processing all  the  samples in the collected dataset $\cup_{k\in G_j} \hat{D}_k^s$. If the computation cannot be finished within $T-\tau_{S,j}^{\text{trans}}$ seconds, we let the satellite  perform computation  for $T-\tau_{S,j}^{\text{trans}}$ seconds and then move back to Step 2-1, i.e.,  the satellite transmits the model and the dataset  to the neighboring satellite that will cover cluster $j$, so that the new satellite can continue training.

These processes in Steps 2-1 and 2-2 are repeated for $N_j$ times, which is written as 
\begin{equation}\label{eq:NJ}
N_j = \left \lfloor  \frac{m_S \sum_{k\in G_j}\alpha_k|D_k|}{(T- \tau_{S,j}^{\text{trans}})f_{S,j}} \right \rfloor.
\end{equation}
Here,  $f_{S,j}$ (in $\text{cycles}/\text{sec}$) is the CPU frequency of satellites focusing on cluster $j$ and $m_S$ (in $\text{cycles}/\text{sample}$) is the number of CPU cycles for the satellite to process one data sample. The term  $m_S \sum_{k\in G_j}\alpha_k|D_k|$ (in $\text{cycles}$) in the numerator denotes the total amount of CPU cycles to finish one local epoch using the collected dataset at the satellite.  The denominator  denotes the number of CPU cycles that each satellite can run within the available time duration $T- \tau_{S,j}^{\text{trans}}$. After repeating this   procedure for $N_j$ times,  the ($N_j+1$)-th satellite 
 (i.e., the last satellite)  receives the model and data,  finishes the remaining updates, and then directly transmits the model and data to the next satellite for the next global round.  Here, the number of CPU cycles that last satellite need to process becomes  $m_S \sum_{k\in G_j}\alpha_k|D_k| - N_j(T - \tau_{S,j}^{\text{trans}})f_{S,j}$ (in $\text{cycles}$). Overall,  each satellite covering cluster $j$ will participate  in the training process  for $\tau_{S,j}^{\text{local}}$ seconds, where 
\begin{align}\label{eq:local_satellite_time}
\tau_{S,j}^{\text{local}} =
\begin{dcases}  
T ,
\      \text{(The first $N_j$ satellites}),\\
\frac{m_S \sum_{k\in G_j}\alpha_k|D_k| - N_j(T - \tau_{S,j}^{\text{trans}} )f_{S,j}}{f_{S,j}} \\ \ \ + \tau_{S,j}^{\text{trans}},  
  \text{(The last satellite)}
\end{dcases}
\end{align}
holds.  The corresponding energy consumption at each satellite becomes
\begin{align}\label{eq:local_satellite_energy}
E_{S,j}^{\text{local}} =
\begin{dcases}  
\kappa \cdot (T - \tau_{S,j}^{\text{trans}})  f_{S,j}^3+ E_{S,j}^{\text{trans}},
\      \text{(First $N_j$ satellites}),\\
\kappa\cdot\Big(m_S \sum_{k\in G_j}\alpha_k|D_k|- N_j(T - \tau_{S,j}^{\text{trans}} )f_{S,j} \Big)f_{S,j}^2 \\ \ \ + E_{S,j}^{\text{trans}},  \  \text{(The last satellite)}.
\end{dcases}
\end{align}
 For the first $N_j$ satellites,  $E_{S,j}^{\text{local}}$ reflects  the energy consumption for  local computation  ($T- \tau_{S,j}^{\text{trans}}$  seconds) and communication ($\tau_{S,j}^{\text{trans}}$ seconds). For the last satellite, it consists of  energy consumption for processing the remaining data samples and then communicating the model/dataset for  $\tau_{S,j}^{\text{trans}}$ seconds for the next global round.

Finally, based on (\ref{eq:local_satellite_time}), the overall latency for Step 2 at cluster $j$ can be written as
\begin{align}\label{tau_SJREP}
&\tau_{S,j}^{\text{rep}} =  \underbrace{T\cdot N_j}_{\text{Latency for the first $N_j$ satellites}} \\ &+  \underbrace{\frac{m \sum_{k\in G_j}\alpha_k|D_k| - N_j(T - \tau_{S,j}^{\text{trans}} )f_{S,j}}{f_{S,j}}+ \tau_{S,j}^{\text{trans}}}_{\text{Latency for the last satellite}}. 
\end{align}
The above latency model is unique to our work that adopts ISL communications    for transmitting the trained model and the offloaded dataset to the next satellite covering the cluster.

%
%

\subsection{Step 3: Intra-Cluster Model Aggregation}
After Steps 1 and 2 are completed, the satellite that is covering cluster $j$ aggregates the client-side models $\{\mathbf{w}_{C,k}^{r+1}\}_{k\in G_j}$
 and the satellite-side model $\mathbf{w}_{S,j}^{r+1}$    according to
 {\small
\begin{equation}
\bar{\mathbf{w}}_j^{r+1}  = \frac{\big(\sum\limits_{k\in G_j}\alpha_k|D_k|\big)\cdot\mathbf{w}_{S,j}^{r+1}  + \sum\limits_{k\in G_j}(1-\alpha_k)|D_k| \mathbf{w}_{C,k}^{r+1}}{\sum\limits_{k\in G_j}|D_k|}.\label{eq:intra_clsuter}
\end{equation}
}

In this process, each client $k\in G_j$ needs to send the updated model to the satellite via uplink communication.  The uplink communication time $\tau_{C,k}^{\text{agg}}$ for client $k$  to upload the model can be written as
\begin{equation}
\tau_{C,k}^{\text{agg}}= \frac{  S(\mathbf{w})}{b_k\log_2(1+\frac{p_{C,k}|h_{C,k}|^2}{ b_kN_0})},
\end{equation}
where $S(\mathbf{w})$ is the model size in bits, $b_k$ is the bandwidth allocated to client $k$, $p_{C,k}$ is the transmit power of client $k$, and  $h_{C,k}$ is the channel between client $k$ and the satellite covering the cluster. As in   \cite{deng2020ultra},  we model the channel $h_{C,k}$ using large-scale fading, which is dominant compared to the effect of small-scale fading. We let $|h_{C,k}|^2 = d_{C,k}^{-\xi}$, where $d_{C,k}$  denotes the distance between client $k$ and the  satellite covering the corresponding cluster, and $\xi$ is the pathloss exponent.   We note that the system parameters should be selected to make the aggregation delay $\tau_{C,k}^{\text{agg}}$  smaller  than the coverage duration $T$. Otherwise, the satellite is not able to aggregate the models within $T$ seconds. 
The corresponding energy consumption for model transmission at  client $k$ becomes
 \begin{equation}
E^{\text{agg}}_{C,k} = p_{C,k} \cdot \tau^{\text{agg}}_{C,k}.
\end{equation}

	\begin{algorithm}[t]
	\small
  	\caption{\small Proposed Satellite-Assisted FL Algorithm}\label{algo:proposed_final}
		\begin{algorithmic}[1]
		\STATE \textbf{Input: }Initialized model $\mathbf{w}^0$ \STATE\textbf{Output:} Global model $\mathbf{w}^R$ after $R$ global rounds\\	
		\FOR{each client $k=1,2,\dots, K$ \textbf{in parallel}} 
	\STATE  Offload $\hat{D}_k^s$ to the corresponding satellite (Preprocessing step)\\ 
	\ENDFOR

		\FOR{each global round $r=0,1,\dots,R-1$}
		      \FOR{each cluster $j\in\{1,2,\dots,J\}$ \textbf{in parallel}}	
			    \FOR{each client $k\in G_j$ \textbf{in parallel}}	
				    				  \STATE   Client  update to obtain $\mathbf{w}^{r+1}_{C, k}$ according to  (\ref{eq:clientlocaljupdate})    (Step 1)
		                \ENDFOR
		                \STATE{Repeated satellite-side update and data/model transmission via ISL   to obtain $\mathbf{w}^{r+1}_{S, j}$ (Step 2,  parallel with Step 1)}
		                		    		                  \STATE  Construct $\bar{\mathbf{w}}_{j}^{r+1}$ via intra-cluster aggregation according to (\ref{eq:intra_clsuter}) (Step 3)  
		    \ENDFOR
		                		    		                  \STATE  Global aggregation $\mathbf{w}^{r+1} = \frac{1}{J}\sum_{j=1}^J \bar{\mathbf{w}}_j^{r+1}$ 
 (Step 4)	    \ENDFOR
	\end{algorithmic}
\end{algorithm}
\setlength{\textfloatsep}{5pt}

\vspace{-3mm}

\subsection{Step 4: Global Model Aggregation} \label{algo_B:aggregation}
After the intra-cluster model aggregation process in Step 3, $J$ different satellites obtain different models $\bar{\mathbf{w}}_1^{r+1},  \bar{\mathbf{w}}_2^{r+1},\dots,\bar{\mathbf{w}}_J^{r+1}$ corresponding to each cluster. Following the process of   \cite{Razmi1,  Elmahallawy3}, these models can be  aggregated according to 
\begin{equation}\label{eq:global_AGG}
\mathbf{w}^{r+1} = \frac{1}{J}\sum_{j=1}^J \bar{\mathbf{w}}_j^{r+1}
\end{equation}
at the ground station with the aid of ISL communications. 
 We assume a fixed delay $\tau_j^{\text{glob}}$ for this process for each cluster $j$, i.e.,  $\tau_j^{\text{glob}}$ denotes the delay for the satellite covering cluster $j$ to transmit the aggregated model to the ground station. After the global model $\mathbf{w}^{r+1}$ is constructed by aggregating the models of all clusters according to (\ref{eq:global_AGG}),  at the beginning of next round\footnote{In the next round, each cluster could be  covered by another orbit  different from the one in the previous round. In this case, the  satellite in the previous round can transmit its data/model to the new satellite through inter-plane ISL.}, $\mathbf{w}^{r+1}$ is sent to all clients in the system  which requires  additional latency of $\tau_j^{\text{sync}}$ for cluster $j$.

Steps 1, 2, 3  and 4 are repeated for $R$ global rounds, with Steps 1 and 2 working in parallel. Algorithm \ref{algo:proposed_final} summarizes the overall process of our satellite-assisted FL methodology.

\section{Convergence Analysis}\label{sec:convergence}

In this section, we analyze the  convergence behavior of our algorithm. Given $\hat{D}_k^c$ and $\hat{D}_k^s$ for all $k\in\mathcal{K}$, we define our objective function as 
\begin{equation}\label{eq:conventional_objective}
F(\mathbf{w}) = \frac{1}{J}\sum_{j=1}^J F_j(\mathbf{w}),
\end{equation}
where 
\begin{align}\label{eq:Fj}
& F_j(\mathbf{w}) =  \frac{\big(\sum\limits_{k\in G_j}|\hat{D}_k^s|\big)\ell_{S,j}(\mathbf{w})  + \sum\limits_{k\in G_j}|\hat{D}_k^c| \ell_{C,k}(\mathbf{w})}{\sum\limits_{k\in G_j}|D_k|} \\&=  \frac{\big(\sum\limits_{k\in G_j}\alpha_k|D_k|\big)
\ell_{S,j}(\mathbf{w})  + \sum\limits_{k\in G_j}(1-\alpha_k)|D_k| \ell_{C,k}(\mathbf{w})}{\sum\limits_{k\in G_j}|D_k|} \nonumber
\end{align}
is the local objective function of cluster $j$ defined as the weighted sum of the client-side objective functions in (\ref{eq:client_objective}) and the satellite-side objective function   in (\ref{eq:satellite_objective}). For analysis, we take the following assumptions that are  adopted in existing FL literature \cite{li2019convergence, chang2023asynchronous, reisizadeh2020fedpaq, ganguly2023multi}. 

\begin{assumption}\label{assum:1}
($L$-smoothness) The satellite-side and client-side objective functions   $\ell_{S,j}(\cdot)$ and  $\ell_{C,k}(\cdot)$ are $L$-smooth functions, i.e.,   $\|\nabla\ell_{S,j}(\mathbf{w}) - \nabla\ell_{S,j}(\mathbf{v}) \|\leq L\|\mathbf{w} - \mathbf{v}\|$ and $\|\nabla\ell_{C,k}(\mathbf{w}) - \nabla\ell_{C,k}(\mathbf{v}) \|\leq L\|\mathbf{w} - \mathbf{v}\|$ for any $\mathbf{w}$, $\mathbf{v}$. 
\end{assumption}

\begin{assumption}\label{assum:2}
(Data variability) The local data variability of each client $k$ is bounded as $\|\nabla\ell(x;\mathbf{w}) - \nabla\ell(x';\mathbf{w}) \| \leq \rho\|x-x'\|$ for any $x,x'\in \hat{D}_k^{c}$.  Similarly, $\|\nabla\ell(x;\mathbf{w}) - \nabla\ell(x';\mathbf{w}) \| \leq \rho\|x-x'\|$  holds  for any $x,x'\in \cup_{k\in G_j} \hat{D}_k^{s}$ for the dataset at the satellite.
\end{assumption}
Note that Assumption \ref{assum:1} guarantees the $L$-smoothness of the global loss function  in (\ref{eq:conventional_objective}). To see the effect of the number of processed data samples at each client,  we derive  the convergence bound  considering local update with mini-batch size  $\lambda_{C,k}^r$ at client $k$ at round $r$. To gain further insights, we also introduce the mini-batch size  $\lambda_{S,j}^r$ at the satellite covering cluster $j$ for analysis.  
Now we state the following theorem which describes the convergence behavior of the algorithm. 
\begin{theorem}
(Convergence bound) Suppose Assumptions \ref{assum:1}, \ref{assum:2} hold. Let $\eta_r\leq  \frac{1}{2L}$, and let $\Gamma_R = \sum_{r=0}^{R-1}\eta_r$ be the sum of learning rates.  Then,  the proposed algorithm guarantees the following convergence bound for the non-convex loss function:
\begin{align}
\frac{1}{\Gamma_R}\sum_{r=0}^{R-1}\eta_r\mathbb{E}[\|\nabla F(\mathbf{w}^r)\|^2]  \leq   
\underbrace{\frac{2(F(\mathbf{w}^0) - F^*)}{\Gamma_R}  + \frac{2L\Omega}{\Gamma_R} \sum_{r=0}^{R-1}\eta_r^2}_{=:U(\bar{\alpha}, \bar{\gamma})}, \label{eq:thm1mainresult}
\end{align}
where  $F^*$ is the minimum value of the objective function in (\ref{eq:conventional_objective}),  $U(\bar{\alpha}, \bar{\gamma})$ is the convergence bound written as a function of $\bar{\alpha}=[\alpha_1,\alpha_2,\dots, \alpha_K]$,  $\bar{\gamma}=[\gamma_1,\gamma_2,\dots, \gamma_K]$,  $\mathbb{E}[\cdot]$ is the expectation over the stochasticity of mini-batch SGD, and 
\begin{align}\label{eq:OMEGA}
&\Omega=\frac{2}{  \sum\limits_{j=1}^J \sum\limits_{k\in G_j}|D_k|}\sum_{r=0}^{R-1}\sum_{j=1}^J\Big(\sum_{k\in G_j}\Big(1-\frac{\lambda_{C,k}^r}{|\hat{D}_k^c|}\Big)  \frac{(|\hat{D}_k^c|-1)\rho}{\lambda_{C,k}^r} \nonumber \\ &\times V_{C,k} +\Big(1-\frac{\lambda_{S,j}^r}{|\cup_{k\in G_j}\hat{D}_k^s|}\Big) \frac{(|\cup_{k\in G_j}\hat{D}_k^s|-1)\rho}{\lambda_{S,j}^r}V_{S,j}   \Big), 
\end{align}
\begin{equation} 
V_{C,k} = \frac{\sum_{x\in \hat{D}_k^c} \Big\| x -  \frac{1}{|\hat{D}_k^c|} \sum_{x'\in \hat{D}_k^c} x'\Big\|^2}{|\hat{D}_k^c|-1}, \label{eq:V_def11}
\end{equation}
\begin{equation} 
V_{S,j} = \frac{\sum_{x\in \cup_{k\in G_j}\hat{D}_k^s} \Big\| x -  \frac{1}{|\cup_{k\in G_j}\hat{D}_k^s|} \sum_{x'\in \cup_{k\in G_j}\hat{D}_k^s} x'\Big\|^2}{|\cup_{k\in G_j}\hat{D}_k^s|-1}. \label{eq:V_def22}
\end{equation}
\end{theorem}
\begin{IEEEproof}
See Appendix A.
\end{IEEEproof}
The term $\Omega$ that appears in the right-hand side of (\ref{eq:thm1mainresult}) captures effect of the number of data samples $\lambda_{C,k}^r$ and  $\lambda_{S,j}^r$ processed at each client and satellite, respectively. As $\lambda_{C,k}^r$ and  $\lambda_{S,j}^r$ increase, we can achieve a smaller $\Omega$, which leads to a tighter convergence bound at a specific global round $r$. This advantage is the cost  of  an increased  training time for processing more data samples.  $V_{C,k}$ and $V_{S,j}$ in (\ref{eq:V_def11}) and (\ref{eq:V_def22}) are the sample variance of the dataset  $|\hat{D}_k^c|$ at client $k$ and the dataset $\cup_{k\in G_j}\hat{D}_k^s$ collected at the satellite of cluster $j$, respectively.  
 The convergence of the algorithm  can be guaranteed by decaying the learning rate to satisfy $\Gamma_{R} = \sum_{r=0}^{R-1}\eta_r \rightarrow \infty$ and $\sum_{r=0}^{R-1}\eta_r^2<\infty$. One specific example is $\eta_r = \frac{\eta_0}{1+r}$, which has been  also adopted in prior works \cite{cho2022towards, basu2019qsparse}. It can be seen that the right-hand side of (\ref{eq:thm1mainresult}) goes to zero as $R$ grows, guaranteeing  convergence to the stationary point of the non-convex ML loss function.  
 
\section{Network Optimization for Satellite-Assisted FL}\label{sec:optimization}
In this section,  we formulate the network optimization problem to minimize the training time. Based on the results established in Section \ref{sec:main},  we start by analyzing the   latency of our satellite-assisted methodology.

 \subsection{Latency Analysis for Satellite-Assisted FL}
 
 We first focus on the latency of a specific cluster $j$.  The running time  at  cluster  $j$ to finish a single global round can be written as follows:
\begin{equation}
\tau_j = \underbrace{\max\{\tau_{C,j}, \tau_{S,j}\}}_{\text{Latency until intra-cluster agg.}} + \  \tau_j^{\text{glob}}. \label{eq:latency_for_clusterj}
\end{equation}
In (\ref{eq:latency_for_clusterj}), the first term $\max\{\tau_{C,j}, \tau_{S,j}\}$ is the delay until  the intra-cluster aggregation process is finished in cluster $j$,  while the last term $ \tau_j^{\text{glob}}$ is the additional delay for global aggregation defined in Section \ref{algo_B:aggregation}. Here,  the first term $\max\{\tau_{C,j}, \tau_{S,j}\}$ is affected by two parameters. First, $\tau_{C,j}$ denotes the delay until all clients in cluster $j$ to finish local updates and send the updated models to the   satellite   for intra-cluster aggregation.  The second variable $ \tau_{S,j}$ is the satellite-side latency, which denotes the delay until the last satellite in cluster $j$ finishes its computation.

Now we observe the individual terms. Specifically, $\tau_{C,j}$ can be written as 
\begin{equation}\label{eq:client_side_lateycn}
\tau_{C,j} = \underbrace{\tau_{j}^{\text{sync}}}_{\text{model sync}} + \underbrace{Y_j }_{\text{client model update and upload}},
\end{equation} 
where $\tau_{j}^{\text{sync}}$ is the model synchronization latency for cluster $j$ as defined in Section \ref{algo_B:aggregation} and  $Y_j$ is the   latency at cluster $j$ to finish client-side local computations and intra-cluster aggregation. Specifically, we have the following result: 
\begin{equation}\label{eq:YJJ}
Y_j=\begin{dcases}
T\cdot N_j + \max_{k\in G_j}\{\tau_{C,k}^{\text{agg}} \} \  \ \text{(Case 1)} \\ 
\max_{k\in G_j}\Big\{\max \Big( T\cdot \Big \lfloor  \frac{ \tau_{C,k'}^{\text{local}}}{T} \Big  \rfloor, \tau_{C,k}^{\text{local}}   \Big)  + \tau_{C,k}^{\text{agg}}\Big\}, \   \ \text{(Case 2)}, \\
T\cdot \Big( \Big \lfloor  \frac{   \tau_{C,k'}^{\text{local}}}{T} \Big  \rfloor +1\Big) +   \max_{k\in G_j}\{\tau_{C,k}^{\text{agg}} \},  
 \  \text{otherwise (Case 3),}
\end{dcases}
\end{equation}
where $k' = \argmax_{k\in G_j}\{  \tau_{C,k}^{\text{local}}\}$. Case  1  corresponds to the case with $\max_{k\in G_j}\{  \tau_{C,k}^{\text{local}}\} \leq T\cdot N_j$, while Case 2 indicates $\max_{k\in G_j}\{  \tau_{C,k}^{\text{local}}\} > T\cdot N_j$ and $\max_{k\in G_j}\{\max ( T\cdot  \lfloor  \frac{ \tau_{C,k'}^{\text{local}}}{T}   \rfloor, \tau_{C,k}^{\text{local}} )  + \tau_{C,k}^{\text{agg}}\} \leq  T\cdot (  \lfloor  \frac{   \tau_{C,k'}^{\text{local}}}{T}   \rfloor +1)$. 
To gain insights into (\ref{eq:YJJ}), it is first important to note that  the models sent from the clients are aggregated at the last satellite of the cluster. Moreover, from (\ref{eq:NJ}), recall that the satellite-side computation at cluster $j$ is finished at  the $(N_j + 1)$-th satellite.  Hence, if all clients finish local computations within $T\cdot N_j$, i.e., if  $\max_{k\in G_j}\{  \tau_{C,k}^{\text{local}}\} \leq T\cdot N_j$ holds, all clients can start transmitting the updated model to the satellite when the $(N_j + 1)$-th satellite starts covering the cluster. This corresponds to case 1 of (\ref{eq:YJJ}). However,  some clients may not finish local updates until the $(N_j + 1)$-th  satellite arrives, i.e., $\max_{k\in G_j}\{ \tau_{C,k}^{\text{local}}\}> T\cdot N_j$. For this case, the delay is affected by   client $k'$ that has the largest computation time in the cluster, and either the  $\Big \lfloor  \frac{ \tau_{C,k'}^{\text{local}}}{T} \Big  \rfloor$-th satellite or the $\Big(\Big \lfloor  \frac{ \tau_{C,k'}^{\text{local}}}{T} \Big  \rfloor + 1\Big)$-th satellite aggregates the models depending on the aggregation time. These scenarios correspond to cases 2 and 3 in (\ref{eq:YJJ}).

On the other hand, the satellite-side latency  $\tau_{S,j}$ can be  written as 
\begin{equation}\label{ddddfdfdddd}
\tau_{S,j} = \underbrace{\tau_{j}^{\text{sync}}}_{\text{model sync}} + \underbrace{\tau_{S,j}^{\text{rep}}}_{\text{Satellite-side model update}},
\end{equation}
where $\tau_{S,j}^{\text{rep}}$ is the satellite-side delay to process all data samples, as defined in  (\ref{tau_SJREP}).

Finally, the completion time for one global FL round can be written as  the maximum of the delays of all clusters. This results in
\begin{align}
&\tau^{\text{round}} = \max_{j\in\mathcal{J}}\{\tau_j\} \\ &\underset{(a)}{=} \max\left\{\max_{j\in\mathcal{J}}\{\tau_{C,j} + \tau_{j}^{\text{glob}}\}, \max_{j\in\mathcal{J}}\{\tau_{S,j} + \tau_{j}^{\text{glob}}\} \right\}\underset{(b)}{=} \\
&\max\left\{\max_{j\in\mathcal{J}}\{\tau_j^{\text{sync}} + Y_j + \tau_{j}^{\text{glob}}\}, \max_{j\in\mathcal{J}}\{\tau_j^{\text{sync}}  + \tau_{S,j}^{\text{rep}} + \tau_{j}^{\text{glob}}\}\right\},\label{mainobjectiiiv}
\end{align}
where $(a)$ comes from (\ref{eq:latency_for_clusterj}) and $(b)$ comes from (\ref{eq:client_side_lateycn}) and (\ref{ddddfdfdddd}).

\vspace{-1mm}

 \subsection{Formulation: Latency Minimization}
Now we formulate the following optimization problem to minimize the latency for one global round:
  \begin{subequations}
\begin{align}
 & \  \min_{\bar{\alpha}, \bar{\gamma},  \bar{f}_{S}, \bar{b}} \tau^{\text{round}}  \label{eq:main_onj}  \\
  \text{subject to:} &   \  0\leq\alpha_k\leq\alpha_k^{\text{max}},  \  \forall k\in\mathcal{K} \label{constraint:alpha}  \\ 
 &   \  0\leq\gamma_k\leq 1-\alpha_k,  \  \forall k\in\mathcal{K}\label{constraint:gamma} \\ 
&    \  0\leq f_{S,j} \leq f_{S}^{\text{max}},   \  \forall j\in\mathcal{J}\label{constraint:f_s} \\ 
&  \   \sum_{k\in G_j}b_k \leq B_j, \ \forall j\in\mathcal{J}\label{constraint:bandwidth}\\
 &  \     E_{C,k}^{\text{local}} + E_{C,k}^{\text{agg}}
 \leq  \delta, \  \forall k\in\mathcal{K}\label{ref:constraint_clientbattery}  \\  
 &  \  E_j^{\text{original}}      -   E_{S,j}^{\text{local}}  + \tau_{S,j}^{\text{local}}\cdot P_j^{\text{sun}}   \geq \psi,  \ \forall j\in\mathcal{J}^{\text{sun}}  \label{ref:constraint_satbattery} \\
 &  \  E_j^{\text{original}}      -   E_{S,j}^{\text{local}}    \geq \psi,  \ \forall j\in\mathcal{J}\setminus\mathcal{J}^{\text{sun}} \label{ref:constraint_satbattery222} \\
 &\  U(\bar{\alpha}, \bar{\gamma})\leq \varepsilon\label{eq:constraint_learning} \\
 & \sum_{k\in G_j}\alpha_k|D_k|=A_j\leq A_j^{\text{max}}, \ \forall j\in\mathcal{J}\label{eq:constraint_comm_group}
 \end{align}
 \end{subequations}
 where $\bar{\alpha}=[\alpha_1,\alpha_2,\dots, \alpha_K]$,  $\bar{\gamma}=[\gamma_1,\gamma_2,\dots, \gamma_K]$, $\bar{f}_S=[f_{S,1}, f_{S,2},\dots, f_{S,J}]$,  
  $\bar{b}=[b_1, b_2,\dots, b_K]$.  In (\ref{eq:main_onj}), we optimize $\bar{\alpha}$  in (\ref{eq:alpha_def}) that describes the 
 portion of data offloading at each client,  $\bar{\gamma}$  in  (\ref{eq:gamma}) which denotes the portion of data processed at each client in one global round, $\bar{f}_S$ describing the CPU frequencies of satellites, and $\bar{b}$, which describes the bandwidth allocated to the clients\footnote{Once the variables are optimized, we can fix the amount of data offloading  until the environment drastically changes (e.g., when the battery charging of satellites is no longer available due to the rotation of the Earth).}.

Constraint (\ref{constraint:alpha}) indicates that $\alpha_k$   is upper bounded by the portion of non-sensitive samples in client $k$, while  (\ref{constraint:gamma}) shows the feasible value  of each  $\gamma_k$. Constraint (\ref{constraint:f_s}) indicates that $f_{S,j}$ is upper bounded by the maximum CPU frequency of the satellite.   The constraint for the bandwidth in each cluster is given by (\ref{constraint:bandwidth}), and the constraint for the energy consumption at each client is provided in (\ref{ref:constraint_clientbattery}).  (\ref{ref:constraint_satbattery})  shows  the battery constraints  of the satellites covering  the clusters in    $\mathcal{J}^{\text{sun}}$, i.e., the clusters that  are facing the sun  and that  make battery charging of the satellites possible. (\ref{ref:constraint_satbattery222}) shows the battery constraints at the satellites covering the remaining  clusters   (i.e., $j\in\mathcal{J}\setminus\mathcal{J}^{\text{sun}}$).   $E^{\text{original}}_j$ is the battery of the satellite at the moment of joining  cluster $j$,  $E_{S,j}^{\text{local}}$ is the energy consumption  defined in (\ref{eq:local_satellite_energy}) and  $\tau_{S,j}^{\text{local}}$ is the time duration for each satellite to participate in the training process as defined in (\ref{eq:local_satellite_time}).  $\tau_{S,j}^{\text{local}} \cdot P_j^{\text{sun}}$ is the amount of energy that could be charged from the sun during $\tau_{S,j}^{\text{local}}$ seconds, where $P_j^{\text{sun}}>0$ denotes the power of the sun near cluster $j$. 
  $\psi$ is the  minimum battery  constraint that should be satisfied at the satellite  when leaving the FL system for future uses.   (\ref{eq:constraint_learning}) is the constraint  for   the convergence bound in (\ref{eq:thm1mainresult}) to guarantee a certain level of learning performance. Finally,   (\ref{eq:constraint_comm_group}) is the communication load constraint during  data offloading in each cluster, where $A_j$ denotes the total amount of data transmitted from the clients to the satellite in cluster $j$.  By limiting the amount of communication to the satellite within each cluster for data offloading, 
we   prevent excessive delay, energy consumption, and communication 	during FL.

\subsection{Solution: Iterative Algorithm}
 The above problem turns out to be a non-convex optimization problem. We first relax this problem by setting $\gamma_k^r = 1 - \alpha_k$, which means that each client processes all data samples that have not been offloaded to the satellite. Based on this strategy, it can be seen that the convergence bound in (\ref{eq:OMEGA})  is minimized given $\{\alpha_k\}_{k \in \mathcal{K}}$, resolving the constraints (\ref{eq:constraint_learning}) and (\ref{constraint:gamma}).   This theoretically guarantees the best convergence bound for our algorithm, as can be seen from (\ref{eq:thm1mainresult}).

 To tackle this non-convex optimization problem, we develop a method based on  block-coordinated descent consisting of $I$ iterations.  Starting from  initial values $\bar{\alpha}^{(0)}, \bar{f}_{S}^{(0)}, \bar{b}^{(0)}$, we obtain $\bar{\alpha}^{(I)},  \bar{f}_{S}^{(I)}, \bar{b}^{(I)}$ after $I$ iterations, where each iteration entails the following three steps.

\textbf{(i) Optimize $\bar{\alpha}$ given $\bar{f}_S$,   $\bar{b}$.}  Our first step is to optimize $\bar{\alpha}$ given other variables fixed:
 \begin{align}\label{eq:convertedProblem2}
  \  \min_{\bar{\alpha}} \tau^{\text{round}}   \ 
  \text{subject to:}   \  (\ref{constraint:alpha}),  
  (\ref{eq:constraint_comm_group}).
 \end{align}
In this first subproblem, we consider  two terms that are affected by $\bar{\alpha}$: $Y_{j}$ and $\tau_{S,j}^{\text{rep}}$.   From (\ref{tau_SJREP}), we   note that   $\tau_{S,j}^{\text{rep}}$  is an increasing function of the total amount of data offloaded within each cluster, i.e., $\sum_{k\in G_j}\alpha_k|D_k|=A_j$, but is  not affected by the  individual $\alpha_k$ values when $A_j$ is given. On the other hand, $Y_j$ in (\ref{eq:YJJ})  is an    increasing function of  each $\alpha_k$.   
To minimize the overall latency, we   first aim to find the best $\alpha_k$ that minimizes   $Y_j$ given $A_j$ fixed for each $j$.  This subproblem for each cluster $j$ can be formulated as follows:
 \begin{align}\label{subproblem_alhpha}
 \min_{\{\alpha_k\}_{k\in G_j}} Y_j  \  
 \text{subject to:}   
  \sum_{k\in G_j}\alpha_k|D_k|=A_j, \     0\leq\alpha_k \leq\alpha_k^{\text{max}}.
 \end{align}
Note that in (\ref{eq:YJJ}), $N_j$ is also a fixed value when   $\sum_{k\in G_j}\alpha_k|D_k|=A_j$ is given. According to  (\ref{eq:YJJ}), there are three different cases we need to consider when analyzing $Y_j$.

\textit{Case 1 in  (\ref{eq:YJJ}).} Given all variables fixed, $Y_j$ is minimized in this first case, i.e., when $\max_{k\in G_j}\{  \tau_{C,k}^{\text{local}}\} \leq T\cdot N_j$ holds. To check whether this condition can be satisfied or not, we need to optimize $\alpha_k$ to minimize $\max_{k\in G_j}\{  \tau_{C,k}^{\text{local}}\}$. This is equivalent to finding the smallest $\nu$ that satisfies $\tau_{C,k}^{\text{local}} \leq \nu$ for all $k\in G_j$. Considering that $\tau_{C,k}^{\text{local}}$ is a decreasing function of $\alpha_k$,  one can find the smallest $\nu$ and the corresponding $\alpha_k$ values that satisfy $\sum_{k\in G_j}\alpha_k|D_k|=A_j$  via bisection search, within the range $\alpha_k \in [0, \alpha_k^{\text{max}}]$.  If   $\max_{k\in G_j}\{  \tau_{C,k}^{\text{local}}\} \leq T\cdot N_j$ holds using the obtained $\alpha_k$ values,  these results become the optimal solution of (\ref{subproblem_alhpha}).

\textit{Case 2 in  (\ref{eq:YJJ}).}  If $\max_{k\in G_j}\{  \tau_{C,k}^{\text{local}}\} \leq T\cdot N_j$  does not hold with the obtained $\alpha_k$ above, we consider Case 2  in  (\ref{eq:YJJ}) with $Y_j = \max_{k\in G_j}\{\max  ( T\cdot   \lfloor  \frac{ \tau_{C,k'}^{\text{local}}}{T}    \rfloor, \tau_{C,k}^{\text{local}}    )  + \tau_{C,k}^{\text{agg}} \}$. Here, for each client $k$, we can find the lower bound $\alpha_k^{\text{min}}$ of $\alpha_k$  that satisfies the second constraint $\max  ( T\cdot \lfloor  \frac{ \tau_{C,k'}^{\text{local}}}{T}    \rfloor, \tau_{C,k}^{\text{local}}    )  + \tau_{C,k}^{\text{agg}} \leq  T\cdot  (  \lfloor  \frac{   \tau_{C,k'}^{\text{local}}}{T}   \rfloor +1)$ for all $k\in G_j$.  If   $\alpha_k^{\text{min}}$ is   within the feasible range $[0, \alpha_k^{\text{max}}]$ and $\sum_{k \in G_j}\alpha_k^{\text{min}}|D_k| \leq A_j$ holds, this means that there exists a solution that satisfies the constraint.  
Hence, to minimize $Y_j$, we run the bisection search to find the minimum $\nu$ that satisfies $\max ( T\cdot \lfloor  \frac{ \tau_{C,k'}^{\text{local}}}{T}   \rfloor, \tau_{C,k}^{\text{local}}   )  + \tau_{C,k}^{\text{agg}}\leq \nu$ for all $k\in G_j$, under the constraints $ \sum_{k\in G_j}\alpha_k|D_k|=A_j$  and $\alpha_k \in [0, \alpha_k^{\text{max}}]$. Here, if the obtained $\alpha_k$  does not satisfy   the condition $\max_{k\in G_j} \{\max  ( T\cdot \Big \lfloor  \frac{ \tau_{C,k'}^{\text{local}}}{T}    \rfloor, \tau_{C,k}^{\text{local}}   )  + \tau_{C,k}^{\text{agg}} \} \leq  T\cdot  (  \lfloor  \frac{   \tau_{C,k'}^{\text{local}}}{T}   \rfloor +1)$, we  consider the last case.

\textit{Case 3 in  (\ref{eq:YJJ}).} From Case 3 in (\ref{eq:YJJ}), it can be seen that $Y_j$ is minimized when $ \max_{k\in G_j}\{\tau_{C,k}^{\text{agg}} \}$ is minimized, requiring the  same approach to find the solution as in   Case 1 above.

Now by adopting the   solution $\{\alpha_k\}_{k\in G_j}$   of (\ref{subproblem_alhpha}) for each $j\in \mathcal{J}$,   the   problem  in (\ref{eq:convertedProblem2}) can be converted to finding the best $A_1, A_2,\dots, A_J$ to minimize the $\tau^{\text{round}}$. We first observe how $A_j$ affects $Y_j$. From   (\ref{eq:YJJ}), it can be seen that $Y_j$ is a decreasing function of $A_j$ within the range $\max_{k\in G_j}\{  \tau_{C,k}^{\text{local}}\} > T\cdot N_j$. Here,   $\tau_{C,k}^{\text{local}}$ is a decreasing function of $A_j$ while $T\cdot N_j$ is an increasing function of $A_j$.  In other words, $Y_j$ is a decreasing function of $A_j$ when $A_{j} \in [0, A_j^{\text{up}}]$, where $A_j^{\text{up}}$ is the value obtained by conducting bisection search to make $\max_{k\in G_j}\{  \tau_{C,k}^{\text{local}}\}$ and $T\cdot N_j$  as close as possible within $A_{j} \in [0, A_j^{\text{max}}]$. Here, $A_j^{\text{max}}$  comes from  the communication load constraint at each cluster in (\ref{eq:constraint_comm_group}). 
Note that if $A_j > A_j^{\text{up}}$,  i.e., if $\max_{k\in G_j}\{  \tau_{C,k}^{\text{local}}\} \leq T\cdot N_j$ holds, $Y_j$ in (\ref{eq:YJJ}) becomes an increasing function of $A_j$. Also recall that $\tau_{S,j}^{\text{rep}}$ is an increasing function of $A_j$.  Hence, from (\ref{mainobjectiiiv}),  the best $A_j$ can be found by   bisection search to make $Y_j$ as close as possible to $\tau_{S,j}^{\text{rep}}$ within the range $A_j\in [0, A_j^{\text{up}}]$. We have the following lemma.

	\begin{algorithm}[t]
		\small
 	\caption{\small Algorithm to obtain   $\bar{\alpha}^*$ of (\ref{eq:convertedProblem2})}\label{algo:alpha} 
		\begin{algorithmic}[1]
		\STATE \textbf{Input:} A small $\epsilon$,  \textbf{Output: } $\bar{\alpha}^*=[\alpha_1^*, \alpha_2^*,\dots, \alpha_K^*]$. \\
	\FOR{each group $j=1,2,\dots, J$}
		\STATE{Obtain $A_j^{\text{up}}$  by conducting bisection search to make $\max_{k\in G_j}\{  \tau_{C,k}^{\text{local}}\}$ and $T\cdot N_j$ as close as possible within $A_{j} \in [0, A_j^{\text{max}}]$, where the solution of (\ref{subproblem_alhpha}) is adopted for calculating $\tau_{C,k}^{\text{local}}$. }
			\STATE{Set $\nu_L = 0$, $\nu_U = A_j^{\text{up}}$.}
			\WHILE{$\nu_U-\nu_L\geq \epsilon$}
			\STATE{Set $A= (\nu_L + \nu_U)/2$, and obtain solution $\{\alpha_k\}_{k\in G_j}$ of  (\ref{subproblem_alhpha})   by letting $A_j=A$. Compute $Y_j$ and $\tau_{S,j}^{\text{rep}}$.}
			\STATE{\textbf{if} $Y_j\geq \tau_{S,j}^{\text{rep}}$,  set $\nu_L = A$.}
			\STATE{\textbf{else} set $\nu_U = A$.}
			\ENDWHILE
			\ENDFOR
			\end{algorithmic}
\end{algorithm}
\setlength{\textfloatsep}{10pt}

  \begin{lemma}
The optimal solution $\bar{\alpha}^*=[\alpha_1^*,\alpha_2^*,\dots, \alpha_K^*]$ of (\ref{eq:convertedProblem2}) can be obtained by  Algorithm 
 \ref{algo:alpha}. 
\end{lemma}
 Intuitively,  as more data samples in cluster $j$ are offloaded to the satellite, the satellite-side computation time increases, which results in increased $\tau_{S,j}^{\text{rep}}$. At the same time,  when $A_j$ becomes too large, $Y_j$ also turns out to be an increasing function of offloaded data. This 
makes   sense because even when the clients finish the local computations quickly, they do not send the updated models until the last satellite covers the cluster. These parts are captured in our analysis.

\textbf{(ii) Optimize $\bar{f}_S$ given $\bar{\alpha}$, $\bar{b}$.}  Given $\bar{\alpha}$ and $\bar{b}$, our second  subproblem can be written as 
\vspace{-1mm}
 \begin{align}\label{eq:convertedProblem1}
  \  \min_{ \bar{f}_{S}} \tau^{\text{round}}   \ 
  \text{subject to:}   \  (\ref{constraint:f_s}),  (\ref{ref:constraint_satbattery}), (\ref{ref:constraint_satbattery222}).
 \end{align}
It can be first seen that $\tau^{\text{round}}$ is a decreasing function of $f_{S,j}$ for all $j\in\mathcal{J}$. Hence, we need to increase $f_{S,j}$ as much as possible while satisfying the  satellite-side computation and battery constraints  in (\ref{constraint:f_s}),  (\ref{ref:constraint_satbattery}), (\ref{ref:constraint_satbattery222}).  To integrate  constraints (\ref{ref:constraint_satbattery}) and (\ref{ref:constraint_satbattery222}), we define $P_j^{\text{charge}} = P_j^{\text{sun}}$ if $j\in\mathcal{J}^{\text{sun}}$, and $P_j^{\text{charge}} = 0$, otherwise. Hence, the constraints (\ref{ref:constraint_satbattery}), (\ref{ref:constraint_satbattery222}) can be converted to the following single constraint $E_j^{\text{original}}      -   E_{S,j}^{\text{local}}  + \tau_{S,j}^{\text{local}}\cdot P_j^{\text{charge}}   \geq \psi$. We consider the following two different cases.

\textit{Case 1:} $f_{S}^{\text{max}} \geq  \frac{m_S \sum_{k\in G_j}\alpha_k|D_k|}{(T- \tau_{S,j}^{\text{trans}})}$\textbf{.} From the definition of  $N_j$ in (\ref{eq:NJ}), it can be seen that we have $N_j=0$ when the satellite-side CPU frequency is sufficiently large to satisfy $f_{S,j}\geq  \frac{m_S \sum_{k\in G_j}\alpha_k|D_k|}{(T- \tau_{S,j}^{\text{trans}})}$. This is the case where the satellite  finishes the  computation within the coverage time $T$, which is achievable when the maximum CPU frequency of the satellite $f_{S}^{\text{max}}$ is greater  than or equal to $\frac{m_S \sum_{k\in G_j}\alpha_k|D_k|}{(T- \tau_{S,j}^{\text{trans}})}$. 
 In this case, from (\ref{eq:local_satellite_time})
 and (\ref{eq:local_satellite_energy}), we have $\tau_{S,j}^{\text{local}} = \frac{m_S \sum_{k\in G_j}\alpha_k|D_k|}{f_{S,j}} + \tau_{S,j}^{\text{trans}}$  and $E_{S,j}^{\text{local}} = \kappa\cdot (m_S \sum_{k\in G_j}\alpha_k|D_k|)f_{S,j}^2+ E_{S,j}^{\text{trans}}$. Here, $\tau_{S,j}^{\text{local}}$  is a decreasing function of $f_{S,j}$ while $E_{S,j}^{\text{local}}$ is an increasing function of $f_{S,j}$. Hence, to minimize the  latency while satisfying  (\ref{constraint:f_s}) and (\ref{ref:constraint_satbattery}), we need to increase $f_{S,j}$ as much as possible until the equality constraint of $E_j^{\text{original}}      -   E_{S,j}^{\text{local}}  + \tau_{S,j}^{\text{local}}\cdot P_j^{\text{charge}}\geq \psi$   is satisfied or until $f_{S,j} = f_{S}^{\text{max}}$ holds, within the range $f_{S,j}\in  [\frac{m_S \sum_{k\in G_j}\alpha_k|D_k|}{(T- \tau_{S,j}^{\text{trans}})}, f_{S}^{\text{max}}]$.  Since  $\tau^{\text{round}}$ and  $E_j^{\text{original}}      -   E_{S,j}^{\text{local}}  + \tau_{S,j}^{\text{local}}\cdot P_j^{\text{charge}}$ are decreasing functions of $f_{S,j}$, this  solution can be obtained via bisection search, which guarantees the 
optimality under the   $f_{S,j}$ range.   Algorithm \ref{algo:ff} describes the optimization procedure for $f_{S}^{\text{max}} \geq  \frac{m_S \sum_{k\in G_j}\alpha_k|D_k|}{(T- \tau_{S,j}^{\text{trans}})}$.

\textit{Case 2: }$f_{S}^{\text{max}} <  \frac{m_S \sum_{k\in G_j}\alpha_k|D_k|}{(T- \tau_{S,j}^{\text{trans}})}$\textbf{.}   In this case, we always have $N_j\geq 1$, which means that there exist satellites that fully participate in FL during the coverage time $T$. From (\ref{eq:local_satellite_time})
 and (\ref{eq:local_satellite_energy}), we let $\tau_{S,j}^{\text{local}}=T$ and $E_{S,j}^{\text{local}} =\kappa (T - \tau_{S,j}^{\text{trans}})  f_{S,j}^3+ E_{S,j}^{\text{trans}}$, where $E_{S,j}^{\text{local}}$ is again an increasing function of $f_{S,j}$.  We can  increase $f_{S,j}$ as much as possible until  $E_j^{\text{original}}      -   E_{S,j}^{\text{local}}  + \tau_{S,j}^{\text{local}}\cdot P_j^{\text{charge}}=  \psi$  holds or  until $f_{S,j}$ reaches $f_{S}^{\text{max}}$, where the solution can be written in a closed-form.
 
   	\begin{algorithm}[t]
		\small
 	\caption{\small Algorithm to obtain    $\bar{f}_S^{*}=[f_{S,1}^{*}, f_{S,2}^{*},\dots, f_{S,J}^{*}]$ of (\ref{eq:convertedProblem1}) when $f_{S}^{\text{max}}\geq  \frac{m_S \sum_{k\in G_j}\alpha_k|D_k|}{(T- \tau_{S,j}^{\text{trans}})}$}\label{algo:ff} 
		\begin{algorithmic}[1]
		\STATE \textbf{Input:} A small $\epsilon$.  \textbf{Output:} $\bar{f}_S^{*}=[f_{S,1}^{*}, f_{S,2}^{*},\dots, f_{S,J}^{*}]$.  \\
	\FOR{each group $j=1,2,\dots, J$}
	\STATE{Set $\nu_L=\frac{m_S \sum_{k\in G_j}\alpha_k|D_k|}{(T- \tau_{S,j}^{\text{trans}})}$, $\nu_U=f_{S}^{\text{max}}$.}
	\WHILE{$\nu_U-\nu_L\geq \epsilon$}
	\STATE{$f_S =(\nu_L + \nu_U)/2$.}
	\STATE{ \textbf{if} $ E_j^{\text{original}}    - \kappa\cdot (m_S \sum_{k\in G_j}\alpha_k|D_k|)f_{S,j}^2- E_{S,j}^{\text{trans}}  +\Big(\frac{m_S \sum_{k\in G_j}\alpha_k|D_k|}{f_{S,j}} + \tau_{S,j}^{\text{trans}}\Big)P_j^{\text{charge}} \geq \psi$, set $\nu_L = f_S$. }
	\STATE{\textbf{else}  set $\nu_U = f_S$. }
	\ENDWHILE
	\STATE{$f_{S,j}^* = f_S$}
		\ENDFOR
 	\end{algorithmic}
\end{algorithm}

\setlength{\textfloatsep}{10pt}
We now state the following lemma.
   \begin{lemma}\label{lemma:111}
If $f_{S}^{\text{max}}\geq  \frac{m_S \sum_{k\in G_j}\alpha_k|D_k|}{(T- \tau_{S,j}^{\text{trans}})}$ holds, the  optimal  $\bar{f}_S^{*}=[f_{S,1}^{*}, f_{S,2}^{*},\dots, f_{S,J}^{*}]$ of (\ref{eq:convertedProblem1}) can be obtained by Algorithm \ref{algo:ff}. Otherwise, i.e., $f_{S}^{\text{max}}< \frac{m_S \sum_{k\in G_j}\alpha_k|D_k|}{(T- \tau_{S,j}^{\text{trans}})}$, we have $f_{S,j}^{*} = 
\min\Big\{f_{S}^{\text{max}}, \sqrt[\leftroot{-2}\uproot{10}3]{\frac{\max\{E^{\text{original}}   - E_{S,j}^{\text{trans}}  + T  \cdot  P_j^{\text{charge}} - \psi, 0\} }{\kappa(T-\tau_{S,j}^{\text{trans}})}} \Big\}.$
  \end{lemma}
Intuitively, if  cluster $j$ is facing the sun, i.e., $P_j^{\text{charge}} = P_j^{\text{sun}}>0$,  the satellite can  utilize more computation power to satisfy the battery constraint in (\ref{eq:local_satellite_energy}), compared to the case when the cluster is not facing the sun i.e., $P_j^{\text{charge}}=0$.  This effect can be  observed from our solution in Lemma \ref{lemma:111}.

 \textbf{(iii) Optimize $\bar{b}$ given $\bar{\alpha}$,  $\bar{f}$.} Our last subproblem can be formulated as 
\begin{align}\label{eq:convertedProblem3}
  \  \min_{\bar{b}} \tau^{\text{round}}   \ 
  \text{subject to:}   \     (\ref{constraint:bandwidth}), (\ref{ref:constraint_clientbattery}).
 \end{align}
 The energy   term $E_{C,k}^{\text{local}} + E_{C,k}^{\text{agg}}
$  in  (\ref{ref:constraint_clientbattery})  turns out to be  a decreasing function of $b_k$. From (\ref{ref:constraint_clientbattery}), one can first obtain  $b_k\geq b_k^{\text{min}}$ via bisection search, where $b_k^{\text{min}}$ is the minimum bandwidth that should be allocated to client $k$ in order to satisfy the   constraint (\ref{ref:constraint_clientbattery}).  Related to latency, $b_k$ only affects the term $\tau_{C,k}^{\text{agg}}$ in (\ref{eq:YJJ}) for each client $k$. Since $\tau_{C,k}^{\text{agg}}$ is a decreasing function of $b_k$, the overall latency is also a decreasing function of $b_k$. Hence,   we need to strategically allocate  $b_k$   to minimize the $Y_j$  until $\sum_{k\in G_j}b_k=B_j$ holds. We consider three different cases according to  (\ref{eq:YJJ}).

	\begin{algorithm}[t]
 		\small
 	\caption{\small Algorithm to obtain    $\bar{b}^*$ of (\ref{eq:convertedProblem3})}\label{algo:bbetata} 
		\begin{algorithmic}[1]
		\STATE \textbf{Input:} A small $\epsilon$. Initialized $\bar{b}=[b_1^{\text{min}}, b_2^{\text{min}},\dots, b_K^{\text{min}}]$,  \STATE \textbf{Output:} $\bar{b}^*=[b_1^*, b_2^*,\dots, b_K^*]$. \\
			\FOR{each group $j=1,2,\dots, J$}	
			\STATE{For each $k\in G_j$, obtain $b_k^{\text{min}}$ that satisfies $E_{C,k}^{\text{local}} + E_{C,k}^{\text{agg}}
=\delta$ via bisection search. }
							\STATE{For each $k\in G_j$, obtain $b_k^{\text{low}}$ that satisfies $\max  ( T\cdot  \lfloor  \frac{ \tau_{C,k'}^{\text{local}}}{T}   \rfloor, \tau_{C,k}^{\text{local}}  )  + \tau_{C,k}^{\text{agg}} =  T\cdot  (   \lfloor  \frac{   \tau_{C,k'}^{\text{local}}}{T}   \rfloor +1)$ via bisection search.}
\STATE{\textbf{If} $\sum_{k\in G_j}b_k^{\text{low}} \leq B_j$ and $\max_{k\in G_j}\{  \tau_{C,k}^{\text{local}}\} \leq T\cdot N_j$, set $X_k=\max ( T\cdot   \lfloor  \frac{\max_{k\in G_j}\{  \tau_{C,k}^{\text{local}}\}}{T}  \rfloor, \tau_{C,k}^{\text{local}}    )$.}	
\STATE{\textbf{else} Set $X_k=0$.}			
					\STATE{Set an appropriate $\nu_L$ and $\nu_U$ for the bisection search.}
		\WHILE{$\sum_{k\in G_j}b_k<(1-\epsilon)B_j$ or $\sum_{k\in G_j}b_k> B_j$}
		\FOR{each client $k\in G_j$}
		\STATE{Obtain $b_k$ that satisfies $ X_k+ \tau_{C,k}^{\text{agg}}= \frac{1}{2}(\nu_L+\nu_U)$   using   bisection search.}
						\STATE{\textbf{if} $X_k=0$ update $b_k\leftarrow\max\{b_k^{\text{min}}, b_k\}$.}
	\STATE{\textbf{else}  update $b_k\leftarrow\max\{b_k^{\text{min}}, b_k^{\text{low}}, b_k\}$.}
		\ENDFOR
		 \STATE{\textbf{if} $\sum_{k\in G_j}b_k<(1-\epsilon)B_j$, set $\nu_U \leftarrow \frac{1}{2}(\nu_L+\nu_U)$}
 		\STATE{\textbf{else} $\nu_L \leftarrow \frac{1}{2}(\nu_L+\nu_U)$}
 		\ENDWHILE
				\ENDFOR
	\end{algorithmic}
\end{algorithm}
\setlength{\textfloatsep}{5pt}

 \begin{table*}[t]
  \centering
 \caption{ Simulation setting.} \label{table:notations}
 \begin{tabular}{c | c||c | c||c | c}
		\toprule  
		\textbf{Parameter} & \textbf{Value}&\textbf{Parameter} & \textbf{Value} & \textbf{Parameter} & \textbf{Value}\\ 
		\midrule
$f_S^{\text{max}}$ & $10^{10}$ Hz &  
$m_S$ & $3\times 10^{7}$ cycles/sample & $p_{S,j}$ & $10$ W \\
$f_{C,k}$ & $[1,3]\times 10^8$ Hz &  
$m_{C,k}$ & $3\times 10^{7}$ cycles/sample & $p_{C,k}$ & $[0.1,0.3]$  W \\
$T$ & 360 sec & $Q_j^{\text{ISL}}$ & $3.125$ Mbps & $\xi$ & $2$\\
$N_0$ & $3.98\times 10^{-21}$ W/Hz  &$\kappa$ &  $10^{-28} \  \text{Joules}\cdot \text{seconds}^2/\text{cycles}^3$ &    $B_j$& $10$ MHz\\
		\bottomrule
	\end{tabular} 
 \vspace{-3mm}
  \end{table*}
\textit{Cases 1 and 3 in (\ref{eq:YJJ}).}  For  the first and third cases, we have $Y_j=T\cdot N_j + \max_{k\in G_j}\{\tau_{C,k}^{\text{agg}} \}$  and $Y_j = T\cdot (  \lfloor  \frac{   \tau_{C,k'}^{\text{local}}}{T}  \rfloor +1) +   \max_{k\in G_j}\{\tau_{C,k}^{\text{agg}} \}$, respectively. Hence, for these cases, we should allocate bandwidth to minimize $ \max_{k\in G_j}\{\tau_{C,k}^{\text{agg}} \}$. It  can be easily seen that  $Y_j$ is minimized when $\tau_{C,k}^{\text{agg}}$  has the same value for all $k\in G_{j}$ given the constraint $\sum_{k\in G_j}b_k \leq B_J$. Hence, we find $b_k$ that satisfies $\tau_{C,k}^{\text{agg}} = \nu$ for all $k \in G_j$, and change $\nu$ during the bisection search until the equality constraint $\sum_{k\in G_j}b_k = B_j$ is satisfied.  By  considering the constraint $b_k\geq b_k^{\text{min}}$ obtained from (\ref{ref:constraint_clientbattery}), the solution can be achieved with Algorithm \ref{algo:bbetata} by setting $X_k=0$ in Line 11.

\textit{Case 2 in (\ref{eq:YJJ}).} Now we consider Case 2 in   (\ref{eq:YJJ}) with   $Y_j = \max_{k\in G_j}\{\max ( T\cdot   \lfloor  \frac{ \tau_{C,k'}^{\text{local}}}{T}   \rfloor, \tau_{C,k}^{\text{local}}  )  + \tau_{C,k}^{\text{agg}}\}$. Given $\tau_{C,k}^{\text{local}}$, we can find the lower bound for each $b_k\geq b_k^{\text{low}}$ via bisection search that satisfies the second constraint $\max ( T\cdot  \lfloor  \frac{ \tau_{C,k'}^{\text{local}}}{T}   \rfloor, \tau_{C,k}^{\text{local}}  )  + \tau_{C,k}^{\text{agg}} \leq  T\cdot(  \lfloor  \frac{   \tau_{C,k'}^{\text{local}}}{T}    \rfloor +1)$ for all $k\in G_j$.
 If $\sum_{k}b_{k}^{\text{low}} \leq B_j$ holds, we can  use bisection search to allocate bandwidth to make the value $\max ( T\cdot   \lfloor  \frac{ \tau_{C,k'}^{\text{local}}}{T}   \rfloor, \tau_{C,k}^{\text{local}}    )  + \tau_{C,k}^{\text{agg}}$  same for all  $k\in B_j$, while satisfying $b_k\geq b_k^{\text{low}}$ and $\sum_{k\in G_j}b_{k} \leq B_j$. Note that we can optimize $b_k$ with this form only when $\max_{k\in G_j}\{  \tau_{C,k}^{\text{local}}\} \leq T\cdot N_j$ holds. Otherwise, we should consider either Cases 1 or 3 above. We state the following lemma.

     \begin{lemma}
The optimal solution $\bar{b}^*=[b_1^*,b_2^*,\dots, b_K^*]$ of (\ref{eq:convertedProblem3}) can be obtained by running Algorithm \ref{algo:bbetata}.
\end{lemma}

\section{Experimental Results}\label{sec:experiments}

We conduct  experiments using three  benchmark datasets for FL: MNIST, FMNIST  and CIFAR-10. 
We train a convolutional neural network with two convolutional layers and two fully connected layers using MNIST. For FMNIST, we adopt a different model with two convolutional layers and one fully connected layer. Finally, we utilize the VGG-11 model for the CIFAR-10 dataset. We use a PyTorch framework with NVIDIA GeForce RTX 3080Ti GPU  to train  the ML models.

\subsection{Simulation Setup and Baselines}
For simulations, we consider a setup with $K=50$ clients and $J=5$ clusters, each cluster having 10 clients in its region. The altitudes of the LEO satellite orbits are $784$ km from the ground \cite{tang2021computation}. In each global round,  the LEO satellites in the same orbit serially  cover a specific cluster with the squared region of $1200$ m $\times$ $1200$ m  \cite{tang2021computation}. Among 5 clusters, we assume that 3 clusters are facing the sun with $P_j^{\text{sun}}=5$ W.  
 Other  parameter values are shown in Table \ref{table:notations},  
where the satellite-specific parameters are mostly adopted from \cite{rodrigues2023hybrid, Razmi1}. For   MNIST and FMNIST which utilize relatively small models, we set $E^{\text{original}} = 500$ J, $\psi=100$ J. For  CIFAR-10 we set   $E^{\text{original}} = 2000$ J, $\psi=250$ J for training a larger model.   The effect of other $E^{\text{original}}$ values is also studied    in Section \ref{ex:subsec_CCC}.  We set $\alpha_k^{\text{max}}=0.8$ for all users.

\begin{figure}[t]
    \centering
 \begin{subfigure}[b]{0.24\textwidth}
         \centering
         \includegraphics[width=\textwidth]{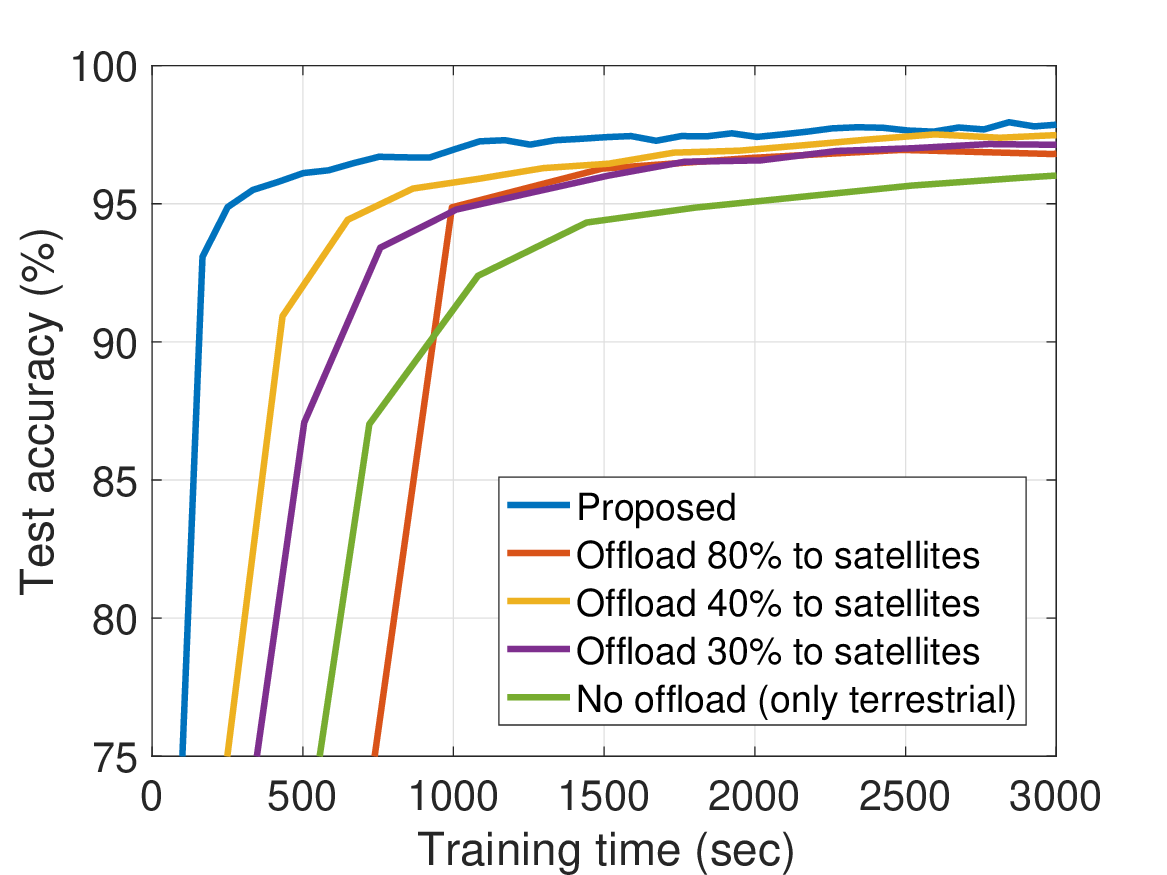}
       \caption{MNIST, IID} 
 \end{subfigure}  
     \begin{subfigure}[b]{0.24\textwidth}
         \centering
         \includegraphics[width=\textwidth]         {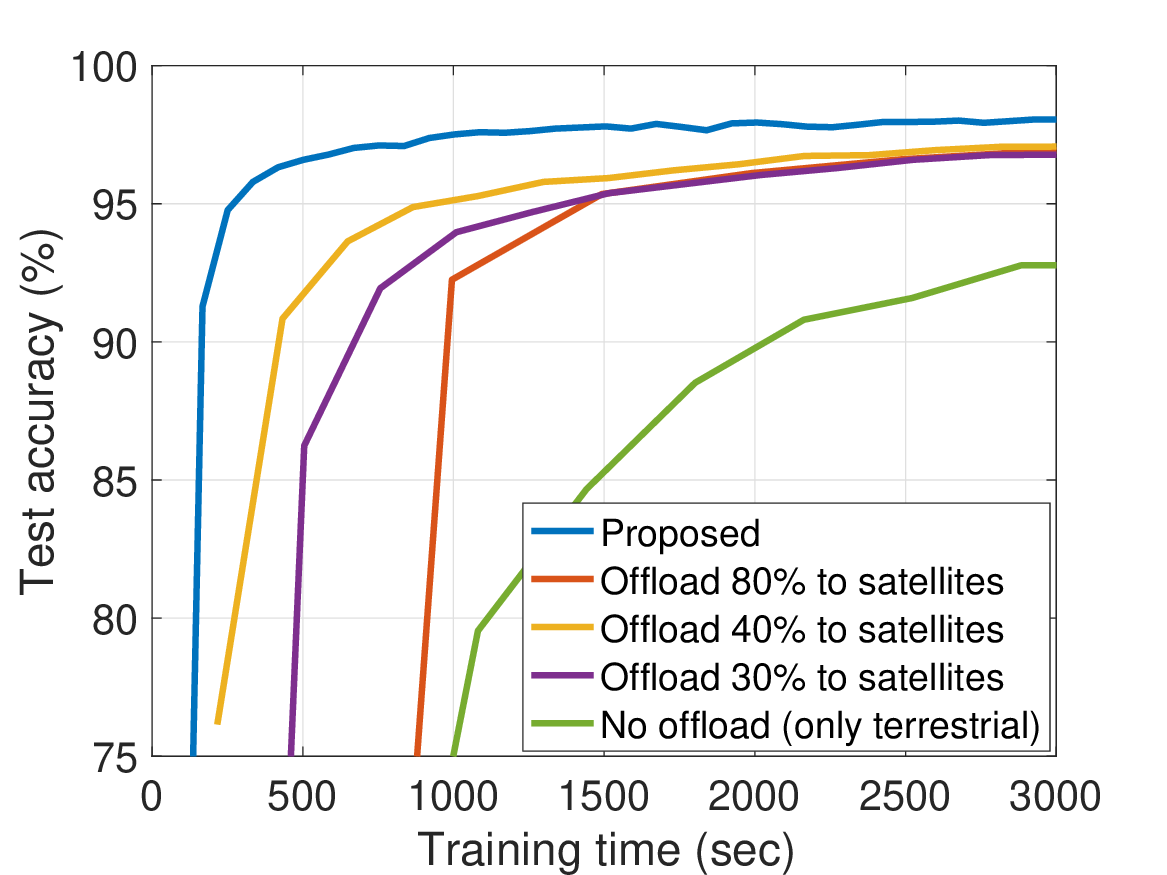}
       \caption{MNIST, Non-IID} 
 \end{subfigure}  
    \begin{subfigure}[b]{0.24\textwidth}
         \centering
         \includegraphics[width=\textwidth]{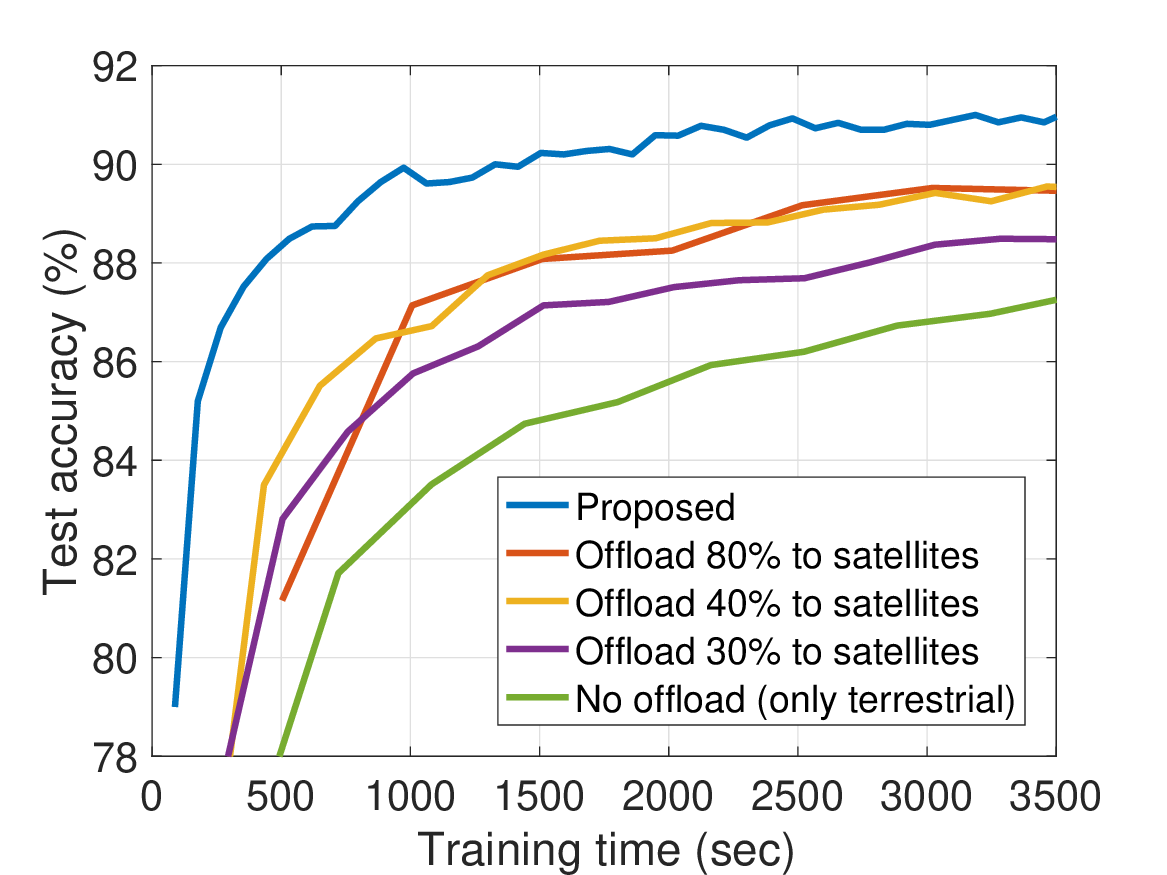}
       \caption{FMNIST, IID} 
  \end{subfigure}
    \begin{subfigure}[b]{0.24\textwidth}
         \centering
         \includegraphics[width=\textwidth]{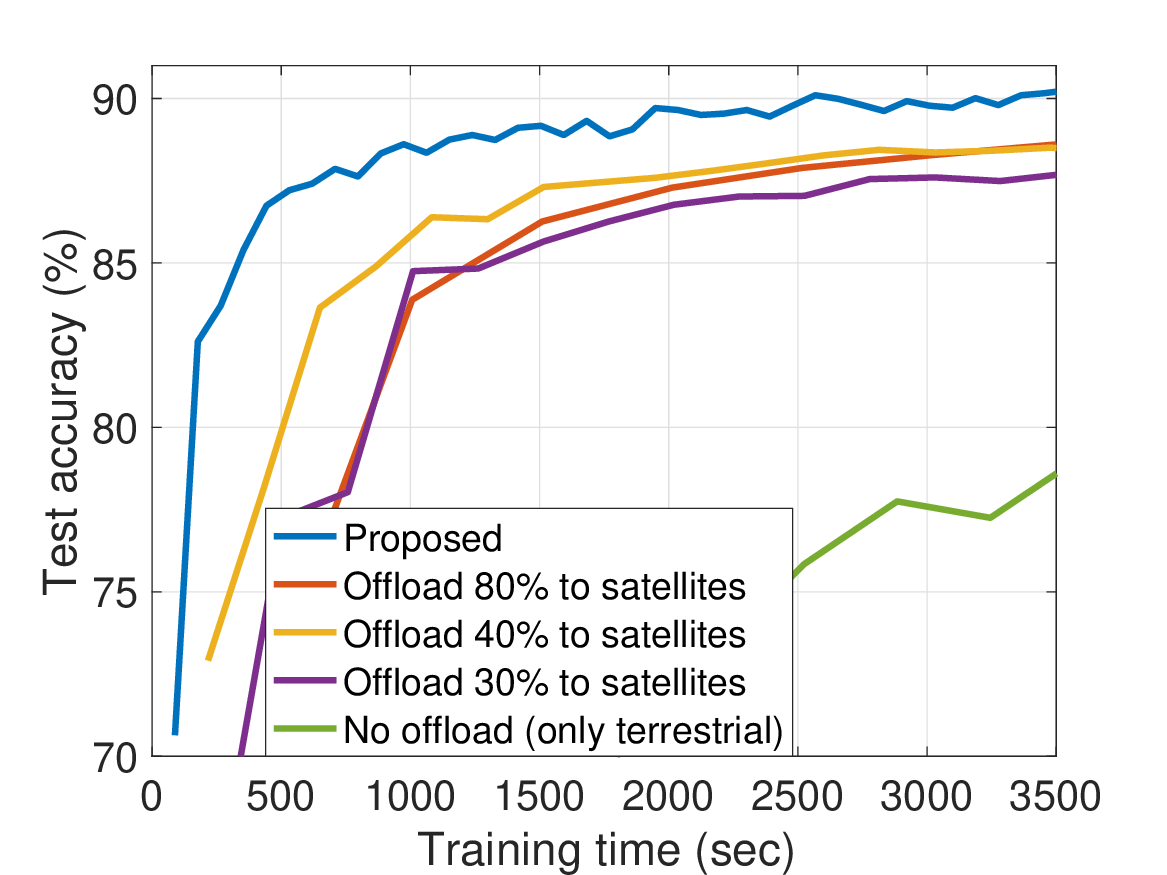}
      \caption{FMNIST, Non-IID} 
  \end{subfigure}
     \begin{subfigure}[b]{0.24\textwidth}
         \centering
         \includegraphics[width=\textwidth]{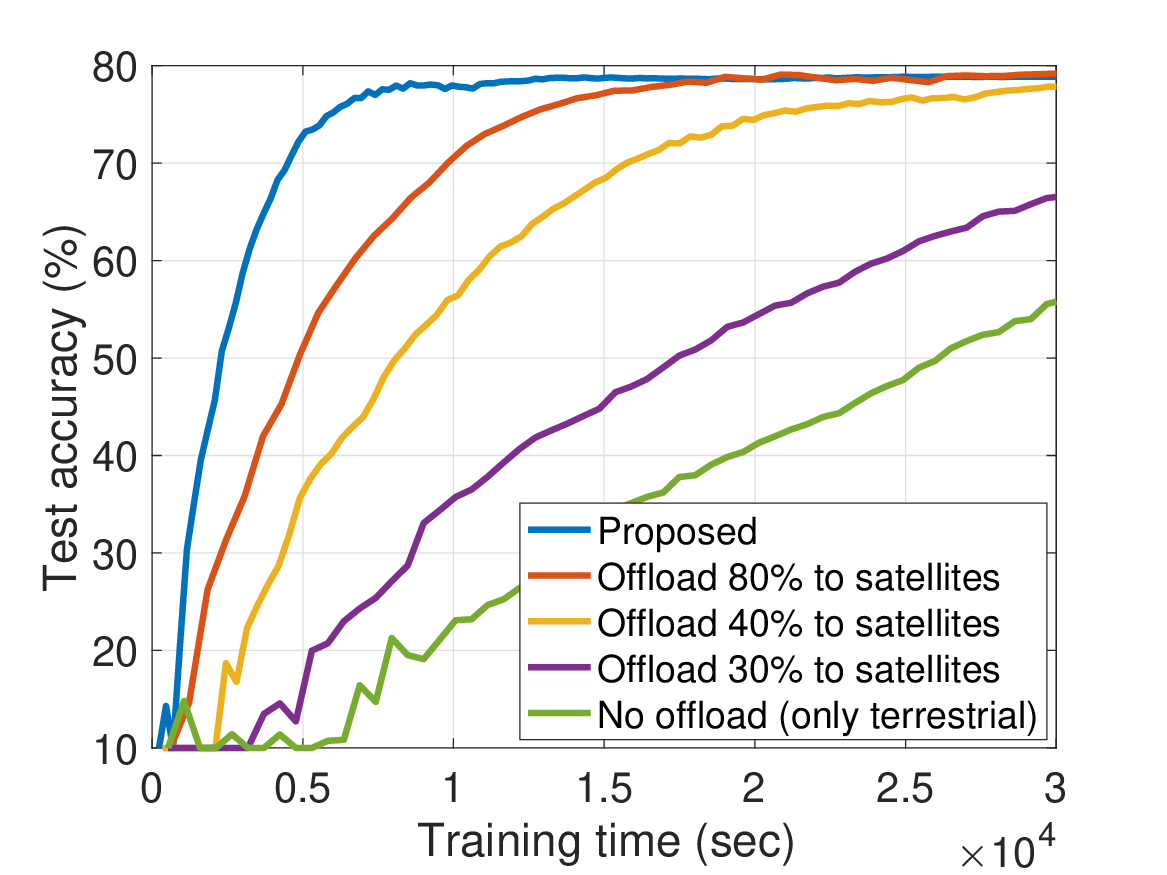}
       \caption{CIFAR-10, IID} 
  \end{subfigure}
     \begin{subfigure}[b]{0.24\textwidth}
         \centering
         \includegraphics[width=\textwidth]{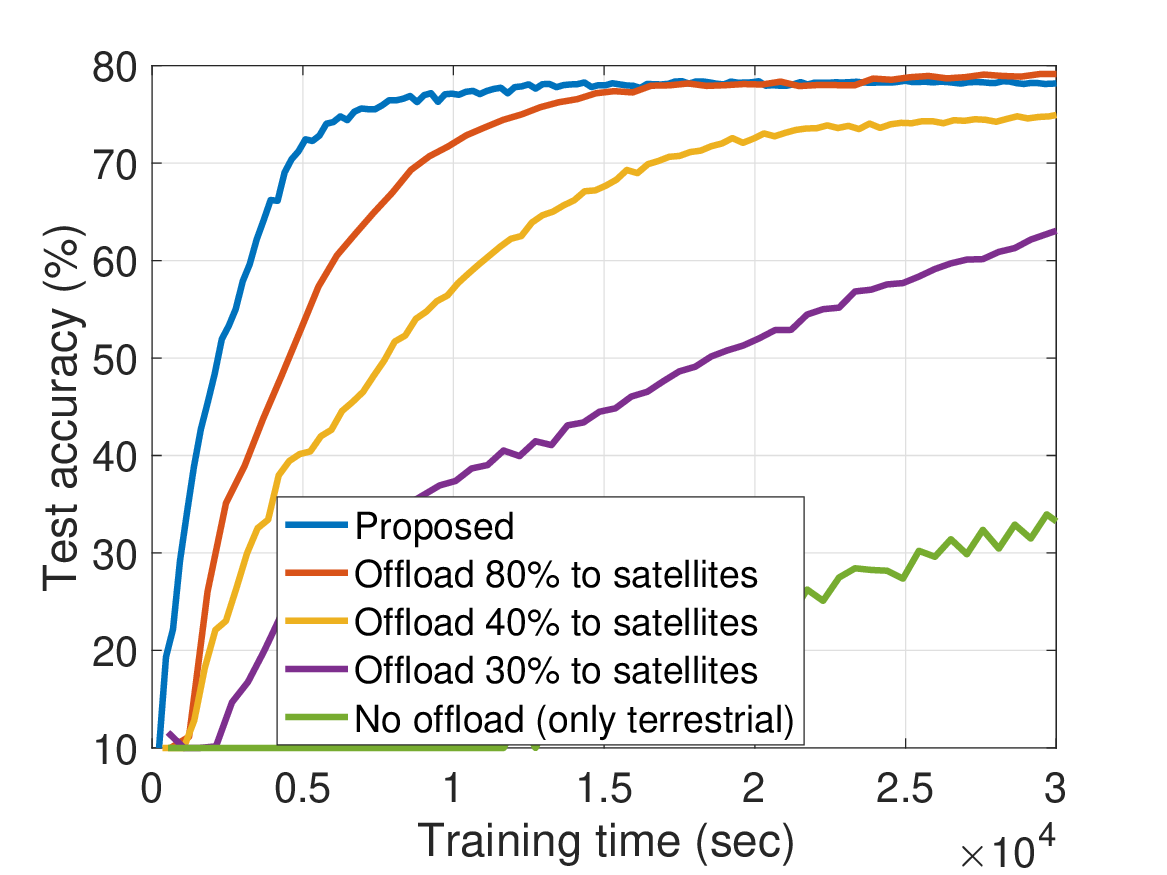}
       \caption{CIFAR-10, Non-IID} 
  \end{subfigure}
     \caption{Test accuracy versus training time. Offloading ratio has been changed from 0 (i.e., only terrestrial) to $80\%$ (i.e., full offloading), considering $\alpha^{\text{max}}=0.8$. }
    \label{fig:exp_main_MNIST}
    \end{figure}

  \begin{figure*}[t]
   \centering 
  \begin{subfigure}[b]{0.29\textwidth}
         \centering
         \includegraphics[width=\textwidth]{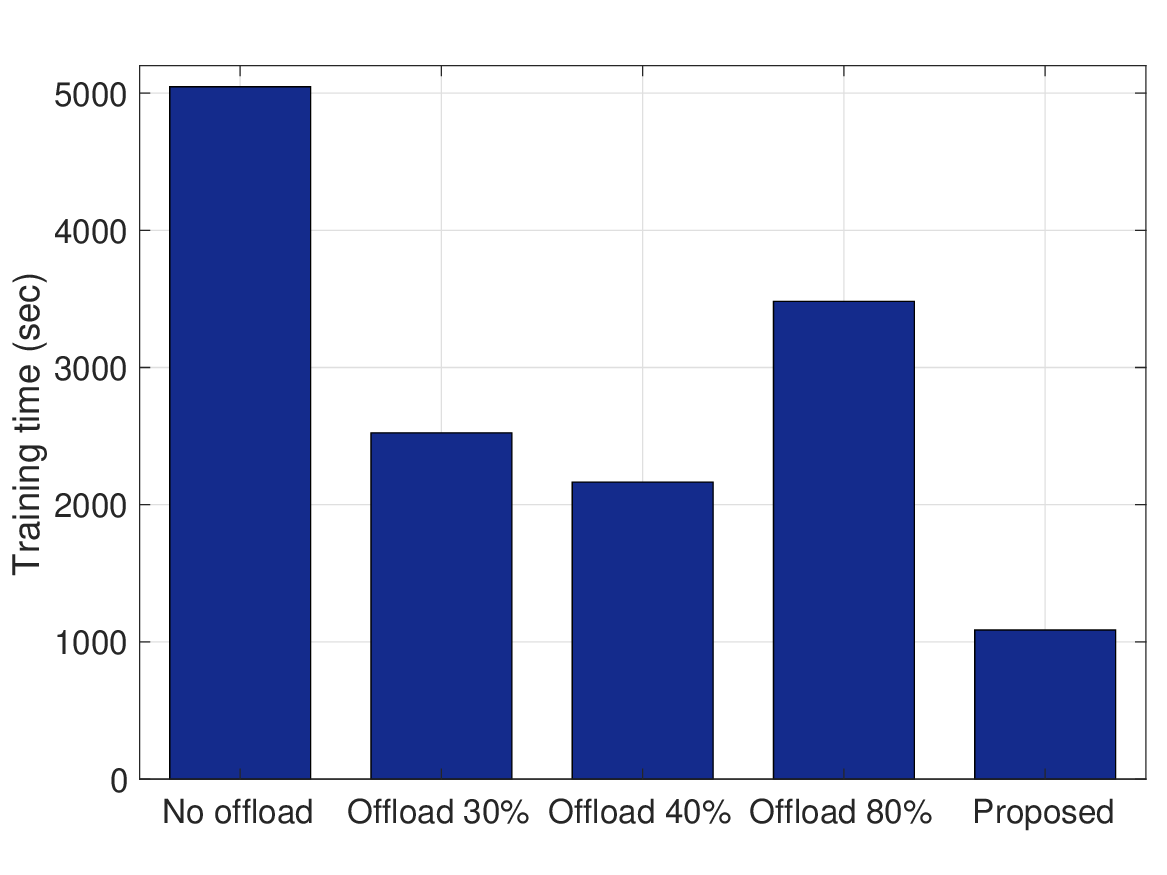}
         \vspace{-6mm}
       \caption{MNIST} 
 \end{subfigure}  
 \begin{subfigure}[b]{0.29\textwidth}
         \centering
         \includegraphics[width=\textwidth]{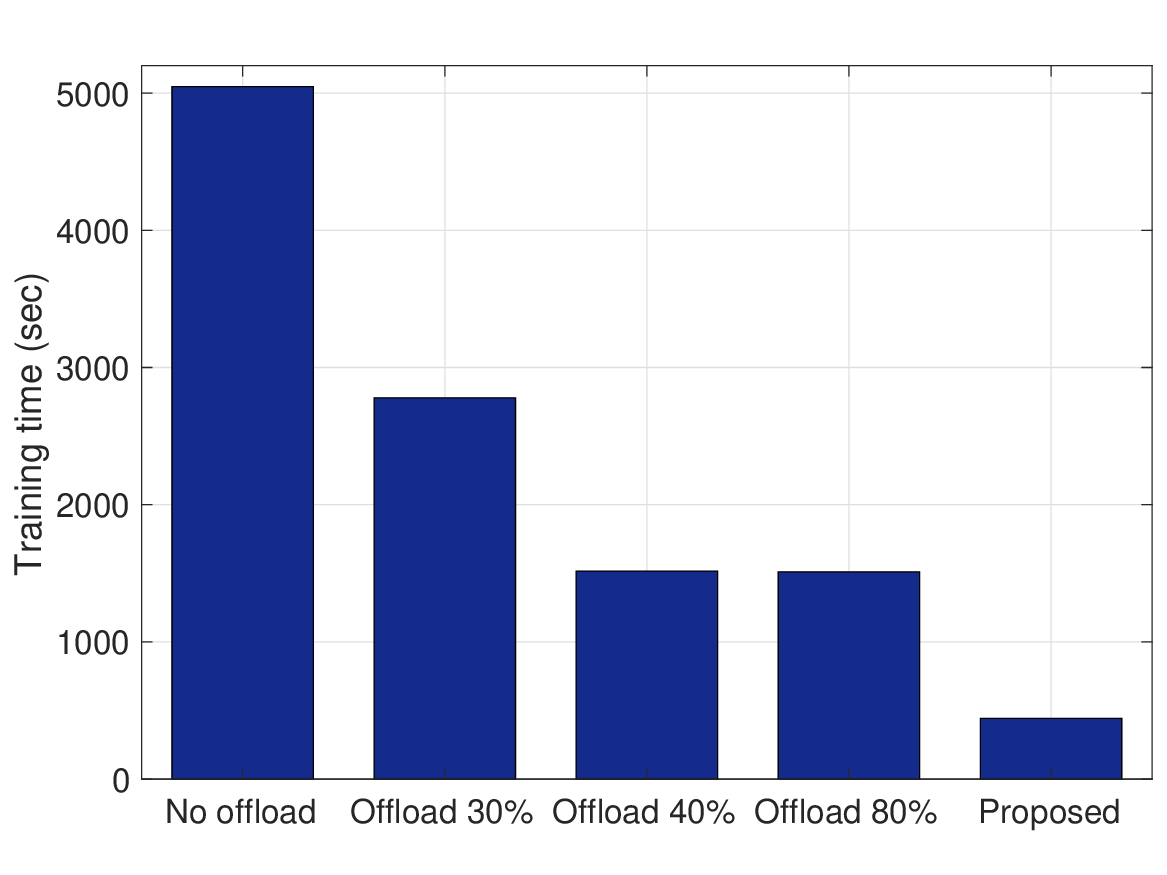}
                  \vspace{-6mm}
       \caption{FMNIST} 
 \end{subfigure}  
   \begin{subfigure}[b]{0.29\textwidth}
         \centering
         \includegraphics[width=\textwidth]{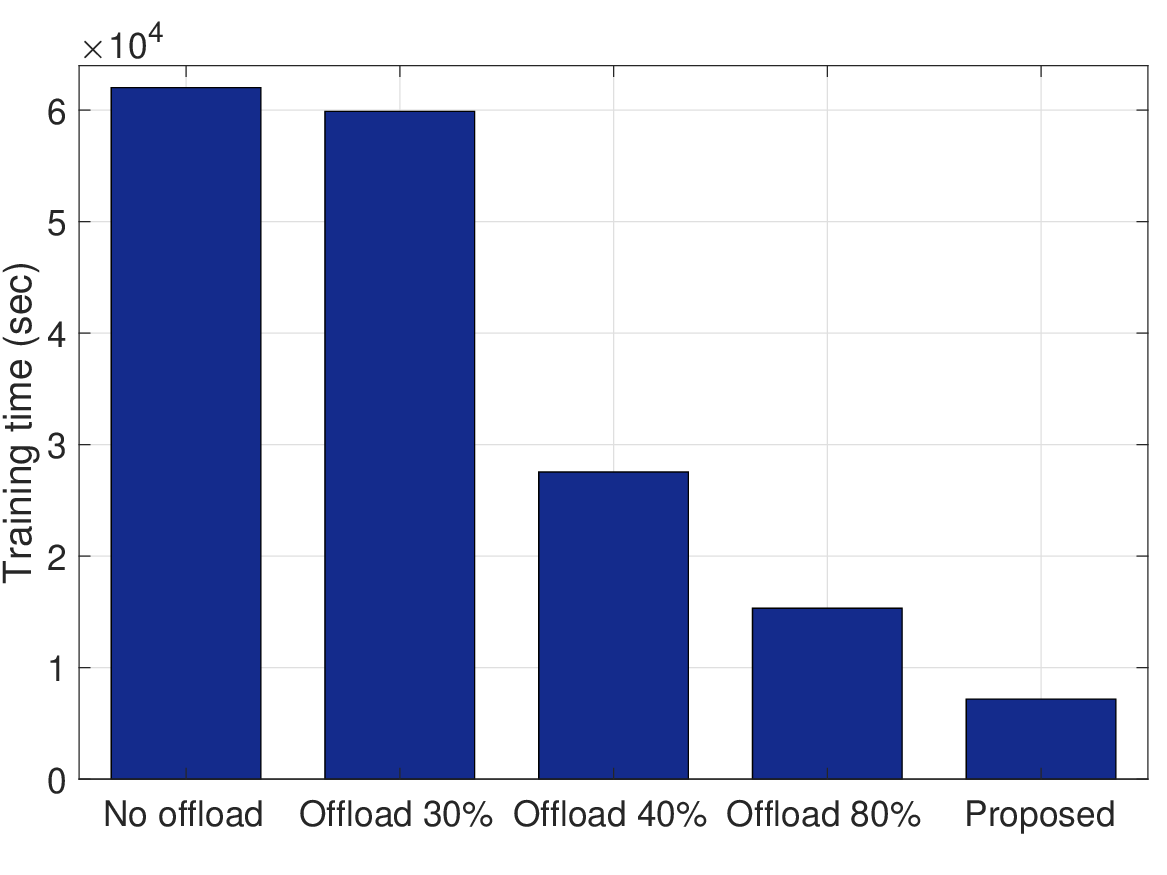}
                  \vspace{-6mm}
       \caption{CIFAR-10} 
  \end{subfigure}
                    \vspace{-1mm}
            \caption{Latency to achieve a certain level of accuracy: 97\% for MNIST, 88\% for FMNIST, 77\% for CIFAR-10.}
    \label{fig:exp_latency}
    \vspace{-3mm}
    \end{figure*}

 \begin{figure*}[t]
  \centering 
  \begin{subfigure}[b]{0.3\textwidth}
         \centering
         \includegraphics[width=\textwidth]{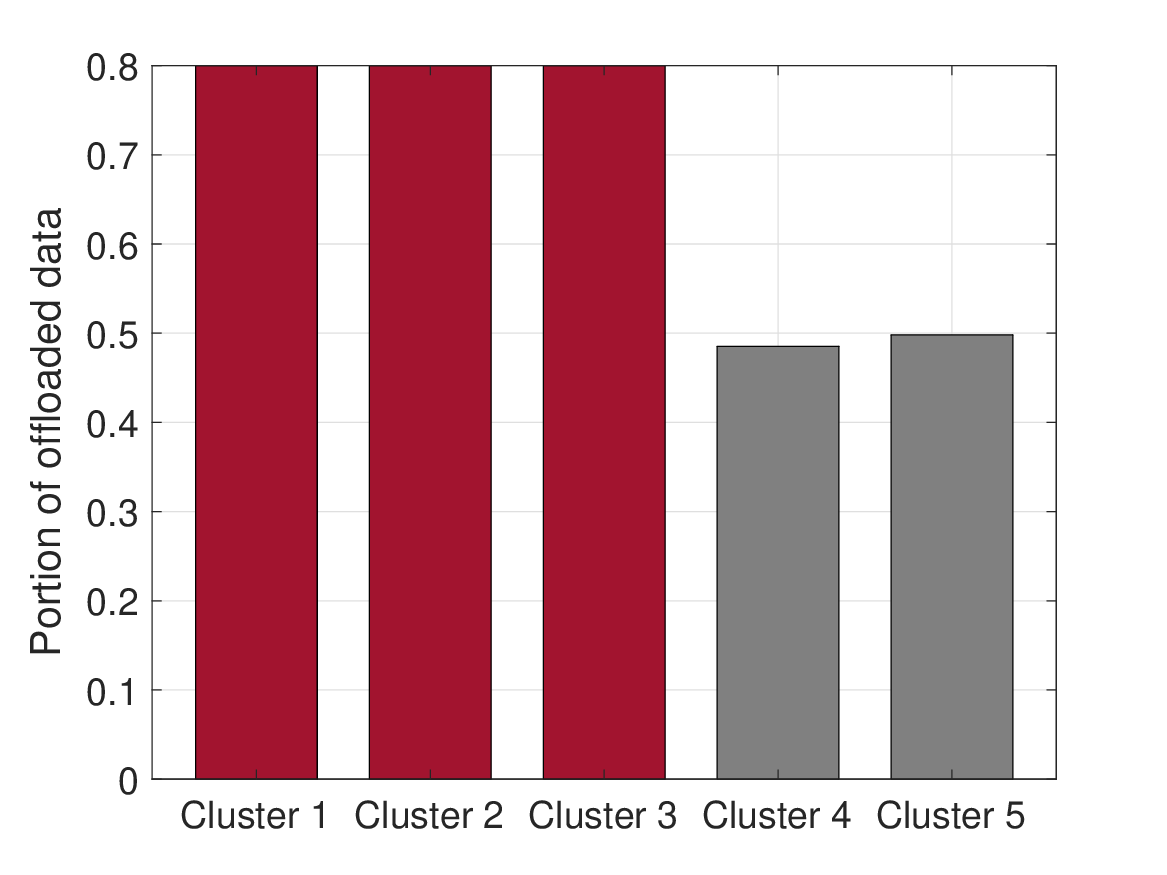}
                \vspace{-6mm}
       \caption{$E^{\text{original}} = 2000$ J} 
 \end{subfigure}  
 \begin{subfigure}[b]{0.3\textwidth}
         \centering
         \includegraphics[width=\textwidth]{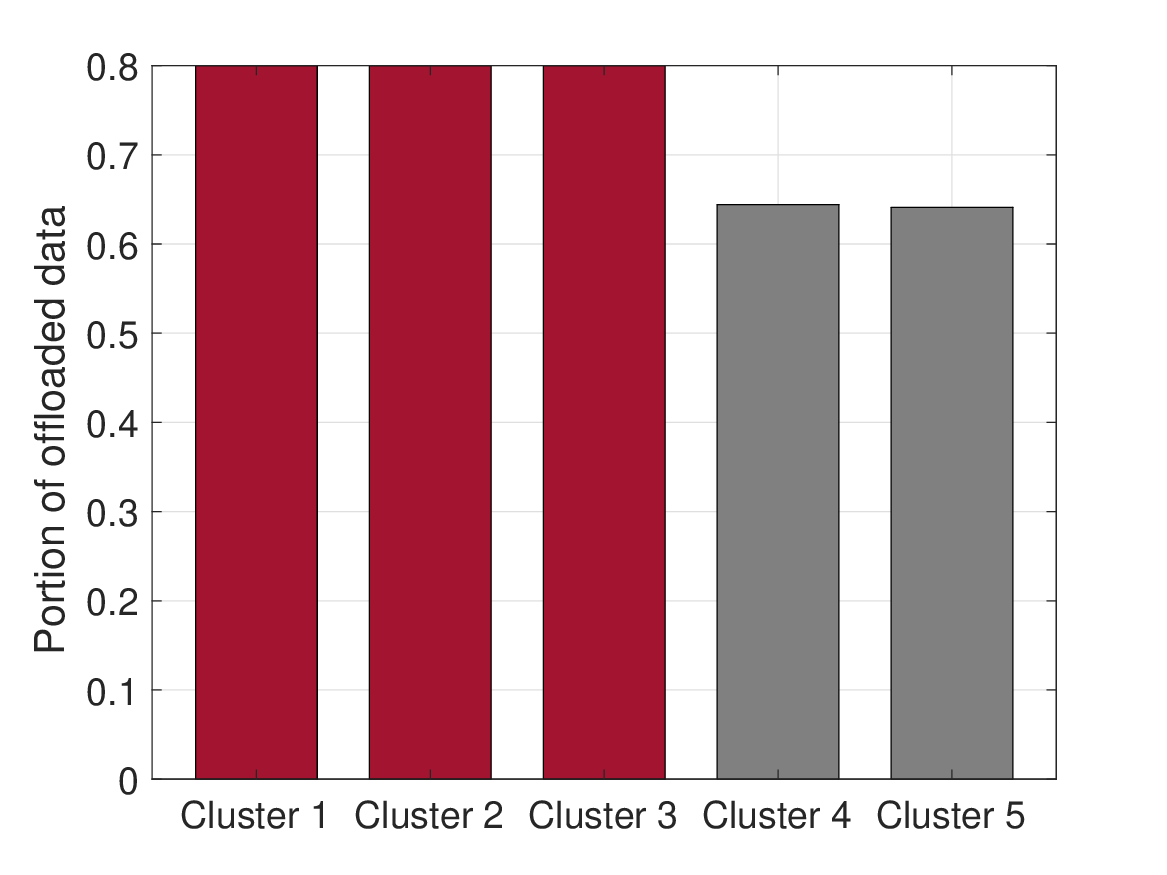}
               \vspace{-6mm}
      \caption{$E^{\text{original}} = 2100$ J} 
 \end{subfigure}  
   \begin{subfigure}[b]{0.3\textwidth}
         \centering
\includegraphics[width=\textwidth]{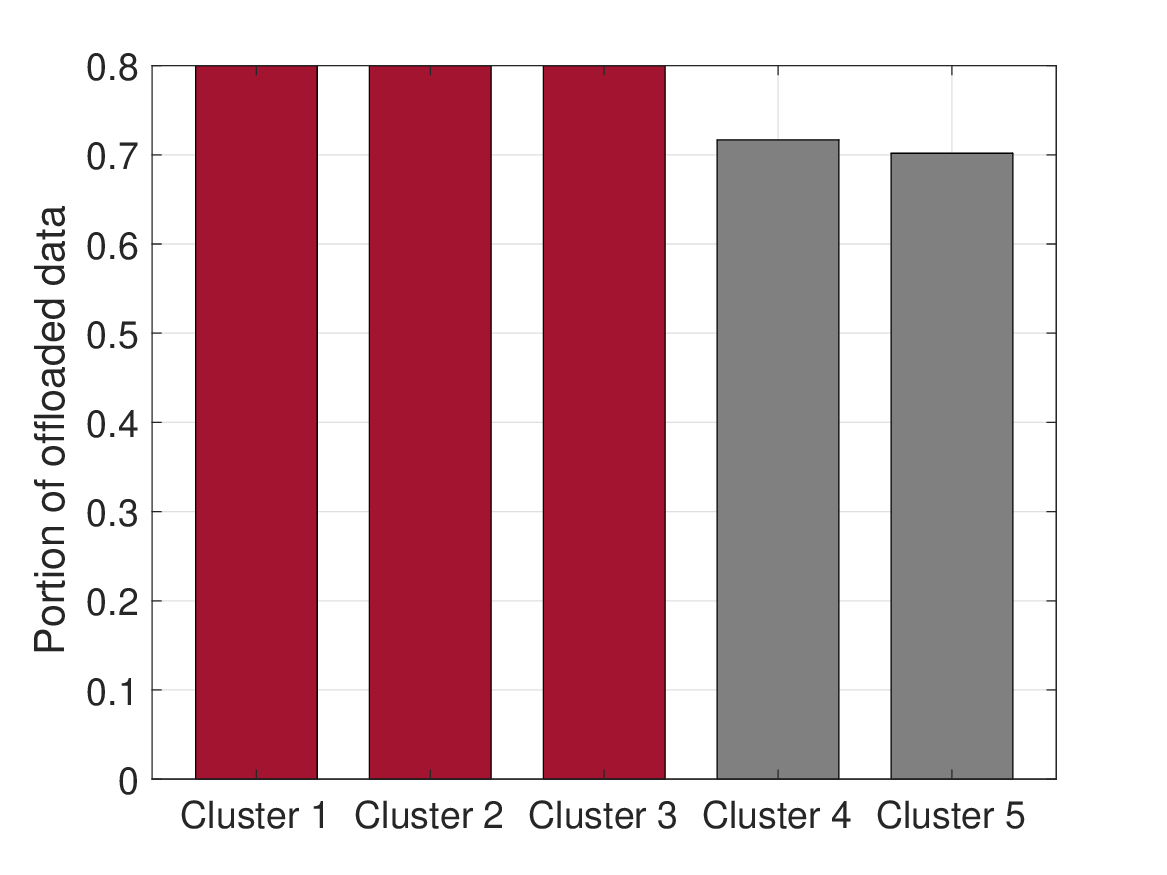}
                \vspace{-6mm}
       \caption{$E^{\text{original}} = 2200$ J} 
  \end{subfigure}
                      \vspace{-1mm}
     \caption{Portion of data offloaded to the satellite within each cluster  in the proposed approach ($\alpha^{\text{max}}=0.8$). Clusters 1, 2, 3 (red bars) are facing the sun, while clusters 4, 5 (gray bars) are not. As the initial battery of the satellite increases, more data samples can be offloaded to the satellite while satisfying the battery constraint.}
    \label{fig:solar}
    \vspace{-4mm}
    \end{figure*}

\textbf{FL implementation.} To model different data distributions across users, we  consider both  the IID (independent, identically distributed) setup  where all clients' local datasets have same distributions, and the non-IID setup where each client has a different data distribution with other clients.  In the IID scenario, the training set of dataset is distributed across the users uniformly at random. For the non-IID case, we first sort the training set based on the  labels and divide it into 100 shards  with equal sizes, as done in  \cite{mcmahan2017communication}. Then, we randomly allocate 2  shards to
each client, which introduces class non-IIDness across users. Each client is allocated with $1200$ data samples for MNIST and FMNIST, as each dataset consists of $60,000$ train samples. For the CIFAR-10  which consists of $50,000$ train samples,  we let each client to have $1000$ samples in its local dataset. During the  local update process, each client trains the model using   mini-batch stochastic gradient descent with momentum of 0.9.  When training is finished, the  test accuracy of the constructed global model  is measured using the test samples in each dataset.

\textbf{Baselines.} We consider the following baselines for comparison. To  confirm the advantage of adopting satellites, we first consider the scheme where all data samples are processed at the client-side without offloading them to the satellites. The satellites are utilized to only aggregate the models of the clients in each cluster. This baseline produces the same model with  the commonly adopted FL strategy for terrestrial networks  \cite{mcmahan2017communication}. Secondly, we let clients to offload all the non-sensitive samples to the satellite, i.e., $\alpha_k=\alpha_k^{\text{max}}$. This baseline maximizes the satellite-side computation/communication burdens for model update and ISL transmission. Finally, we vary the portion of data samples  offloaded from the  devices to   satellites,  and compare the results  with ours.   These   baselines are considered to see the performance of arbitrary data offloading strategies   that are not tailored  to  our delay model.         
 For a fair comparison, we optimized the variables of all
baselines to minimize  latency while satisfying the optimization constraints. 
 Specifically, for the terrestrial-only baseline,  we optimized the user bandwidth   to minimize   latency,     where the latency model of this scheme can be simply obtained by inserting $\alpha_k=0$ to (\ref{mainobjectiiiv}) for all $k=1,2,\dots,K$. For the satellite-based baselines,  we  also optimize the CPU frequency of the satellite to satisfy the satellite-side battery constraints in  (\ref{ref:constraint_satbattery}) and (\ref{ref:constraint_satbattery222}), depending on whether the satellite is facing the sun or not. We  implement  all schemes  including ours using the well-known   FedAvg algorithm \cite{mcmahan2017communication} for a fair comparison.

  \subsection{FL Performance and Latency}\label{ex:subsec_BBB}

Fig. \ref{fig:exp_main_MNIST} shows our main experimental results, comparing    the test accuracies of different schemes as a function of training time.  We make the following  key  observations. First, the scheme that utilizes only the client-side computations without any data offloading to satellites, achieves the worst performance.  One trivial reason is that  this scheme  ignores the satellite-side computation,   not taking the benefit of  parallel  client-satellite computing. Another important factor  is the non-IID data distribution across clients. Without any data offloading, the performance of this baseline becomes extremely limited under the non-IID  setting due to the biased local datasets of clients. Related to this phenomenon, it can be seen that the performance gap between the IID and non-IID cases decreases as more data samples are offloaded to the satellite. The  satellite-side dataset collected from different users can somewhat resolve the issue that arises from the non-IID data distribution. However, offloading too many data samples to the satellite will slow down the training process,  causing latency issues. This can be confirmed by comparing the scheme   that offloads all non-sensitive samples (i.e.,  $\alpha^{\text{max}}=0.8$ portion of local dataset) to the satellite, and the schemes that only offload $30\%$ or $40\%$ of the local datasets.  Even when the maximum computation power of the satellite is large,  collecting large volumes of data at the satellite can cause significant delays due to the   battery constraint at individual satellites.

For our scheme, the average portions of offloaded data   at each client are $56.89\%$, $55.23\%$, and $67.23\%$ for MNIST, FMNIST, and CIFAR-10, respectively.   Compared to the baselines, note that in our scheme, the clients may offload different portions of  samples depending on their computation power, transmit power, and whether the cluster is facing the sun or not. Overall, the results in Fig. \ref{fig:exp_main_MNIST}     shows  that the proposed methodology can provide significant benefits by strategically taking advantage of satellites in FL.

   In Fig. \ref{fig:exp_latency}, we compare the latency of different schemes to achieve a certain level of accuracy. The target accuracies are set to be 97\%, 88\%, 77\% for MNIST, FMNIST, CIFAR-10, respectively, in the IID setup. 
   The overall results are consistent with Fig.   \ref{fig:exp_main_MNIST},  confirming the effectiveness of the proposed FL approach tailored to  ground-to-satellite integrated networks.

\subsection{Ablation Studies and Further Experiments} \label{ex:subsec_CCC}

\textbf{Effect of  satellite-side battery.} To gain insights into the effect of the sun for battery charging, in Fig. \ref{fig:solar}, we compare the portion of offloaded data samples within each cluster of our approach. CIFAR-10 is utilized for training, and the portion of non-sensitive samples are set to be $\alpha^{\text{max}}=0.8$ as in Fig. \ref{fig:exp_main_MNIST}. We have the following  main observations. First, under our simulation setup, the clients in the clusters that are facing the sun (red bars) tend to offload all of the non-sensitive data samples to the corresponding satellite, as the solar-powered satellites have less battery issues when they are facing the sun. On the other hand, if the cluster is not facing the sun (gray bars), the clients offload less data due to the satellite-side battery constraint. As the satellite has more battery in the beginning of training (i.e., as $E^{\text{original}}$ increases), the clients tend to offload more data as the satellites can more easily satisfy the battery constraints. The overall results show  that    the proposed scheme  can strategically optimize  the amount of data to be offloaded and the satellite-side computation power, depending on whether the cluster is facing the sun or not.  
 \begin{figure}[t]
   \centering
   \begin{subfigure}[b]{0.24\textwidth}
         \centering
         \includegraphics[width=\textwidth]{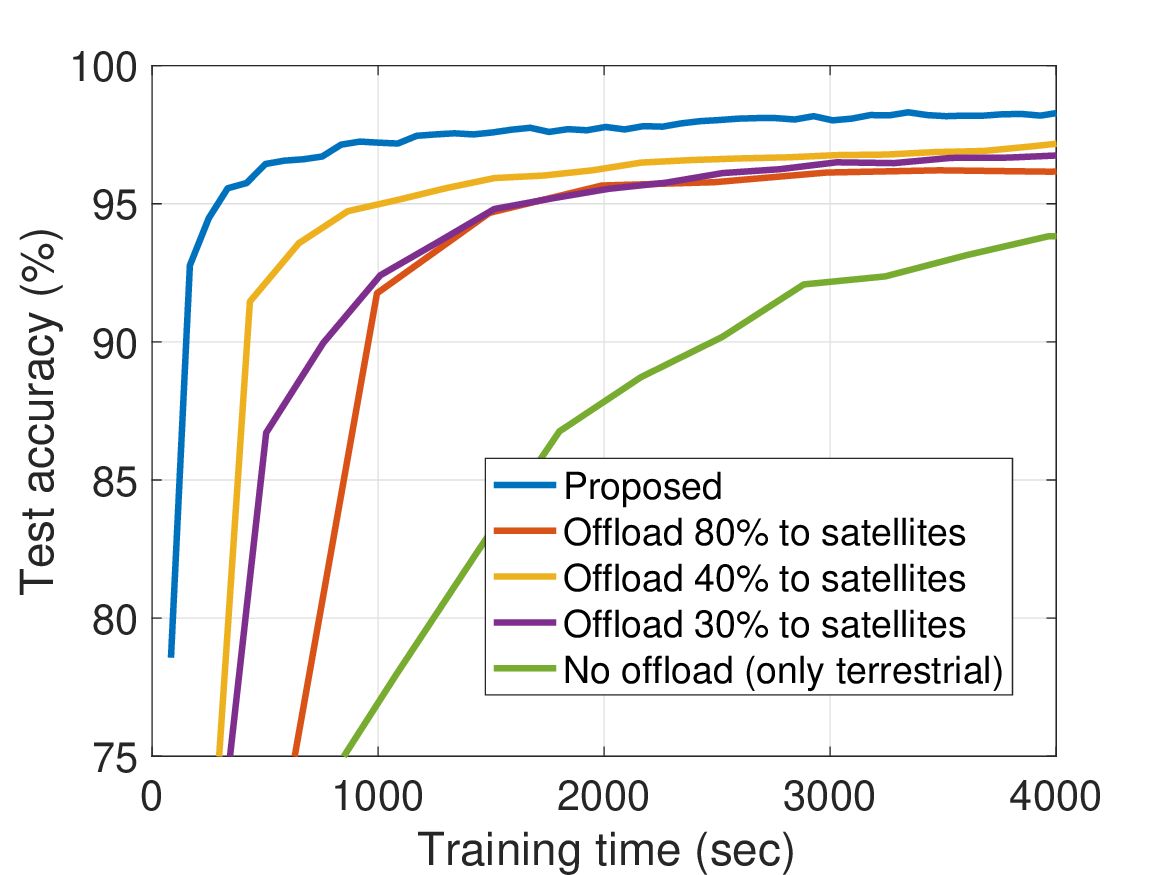}
       \caption{MNIST}
  \end{subfigure}
    \begin{subfigure}[b]{0.24\textwidth}
         \centering
         \includegraphics[width=\textwidth]{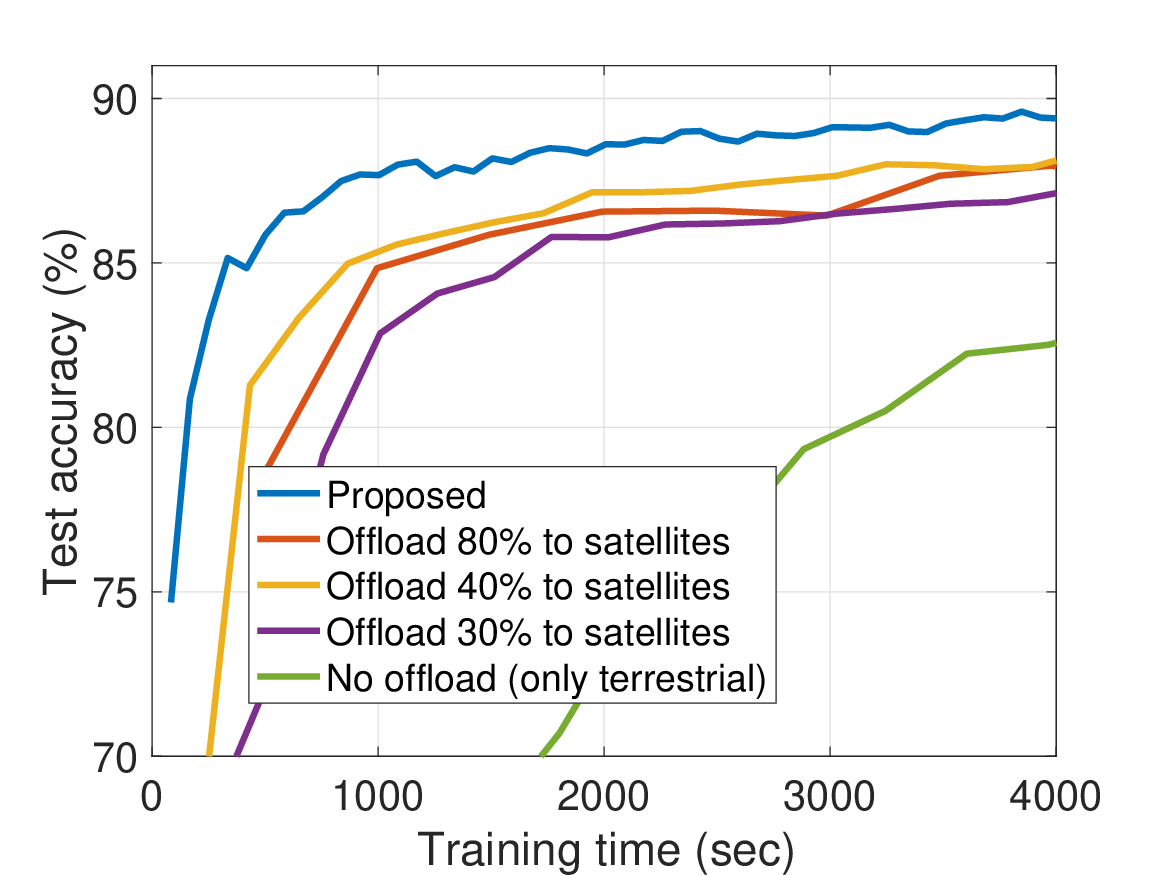}
       \caption{FMNIST}
  \end{subfigure}
     \caption{Compatibility with FedProx.}  
    \label{fig:FedProx}
     \end{figure}
 
\begin{figure}
  \centering
\includegraphics[width=0.32\textwidth]{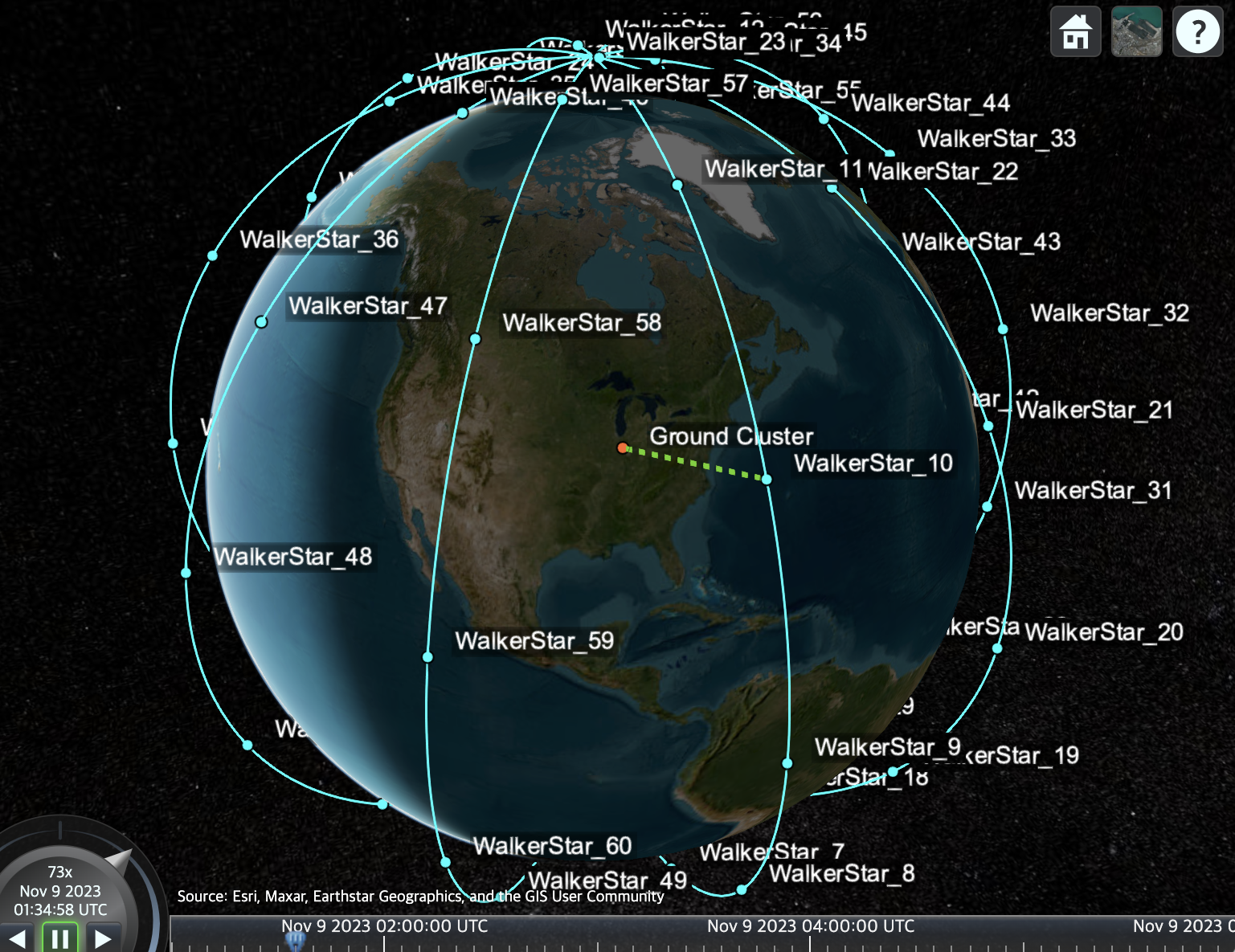}
\caption{Considered  satellite constellation model   for simulating varying coverage times.}\label{fig:Modelinggg}
\end{figure}

\textbf{Compatibility with other FL algorithm.} In Section \ref{ex:subsec_BBB}, we made a fair comparison between our  approach and other baselines by adopting FedAvg to all schemes. To further confirm the advantage of our algorithm, we now use another widely adopted FL algorithm, FedProx \cite{li2020federated_FedProx}, in all baselines and our scheme  for performance comparison.  Fig. \ref{fig:FedProx} shows the result in the non-IID setup. The results are consistent with the ones in previous subsection,  further confirming  the effectiveness and applicability of our satellite-assisted FL methodology.

\textbf{Practical varying coverage times.} Instead of using a fixed the coverage time $T$, we also conduct experiments with varying coverage times using the Walker-Star constellation model in Fig. \ref{fig:Modelinggg}. We specifically use  the \texttt{walkerStar} function \cite{walkerstar} in MATLAB to create a satellite constellation. 
 We consider 50 satellites equally distributed across 5 orbits with altitude of 784 km and inclination of $90^{\circ}$.   The minimum elevation angle to communicate is set to $15^{\circ}$, and  50 clients are distributed in a specific region with latitude of  $40^{\circ}$ N and longitude of $86^{\circ}$ W.   Based on the constructed   constellation and the location of the ground cluster, we    obtained  the coverage times of satellites over the target region using the   \texttt{accessIntervals} function.   Since the coverage time of each satellite can be obtained prior to optimization, we optimize the amount of data offloading considering the average coverage time of future satellites, which is  408 sec.    All remaining setups are the same as in Section \ref{ex:subsec_BBB}.  Fig. \ref{fig:Rebuttal_walker}  shows the results in the IID setup. The results consistently confirm the advantage of the proposed methodology. 

\begin{figure}[t]
  \centering
   \begin{subfigure}[b]{0.24\textwidth}
         \centering
         \includegraphics[width=\textwidth]{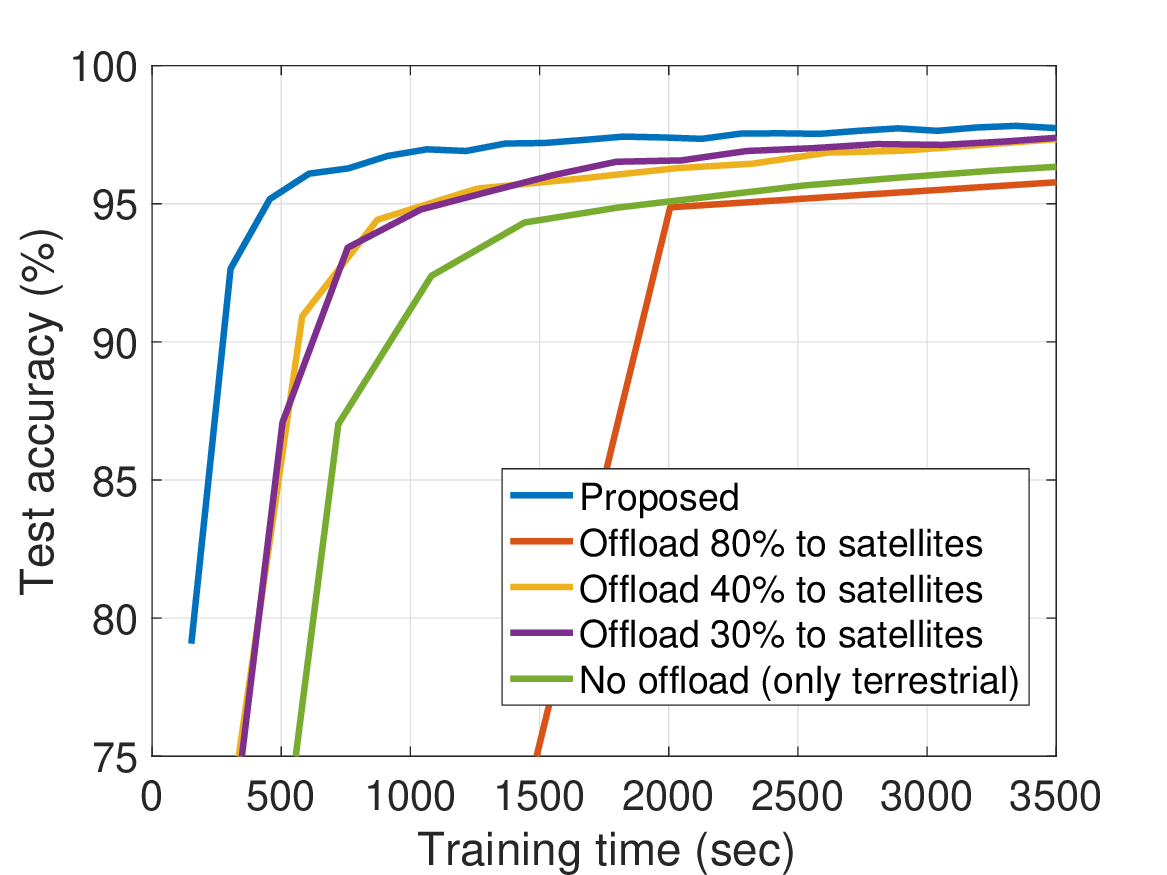}
        \caption{MNIST}
  \end{subfigure}
    \begin{subfigure}[b]{0.24\textwidth}
         \centering
\includegraphics[width=\textwidth]{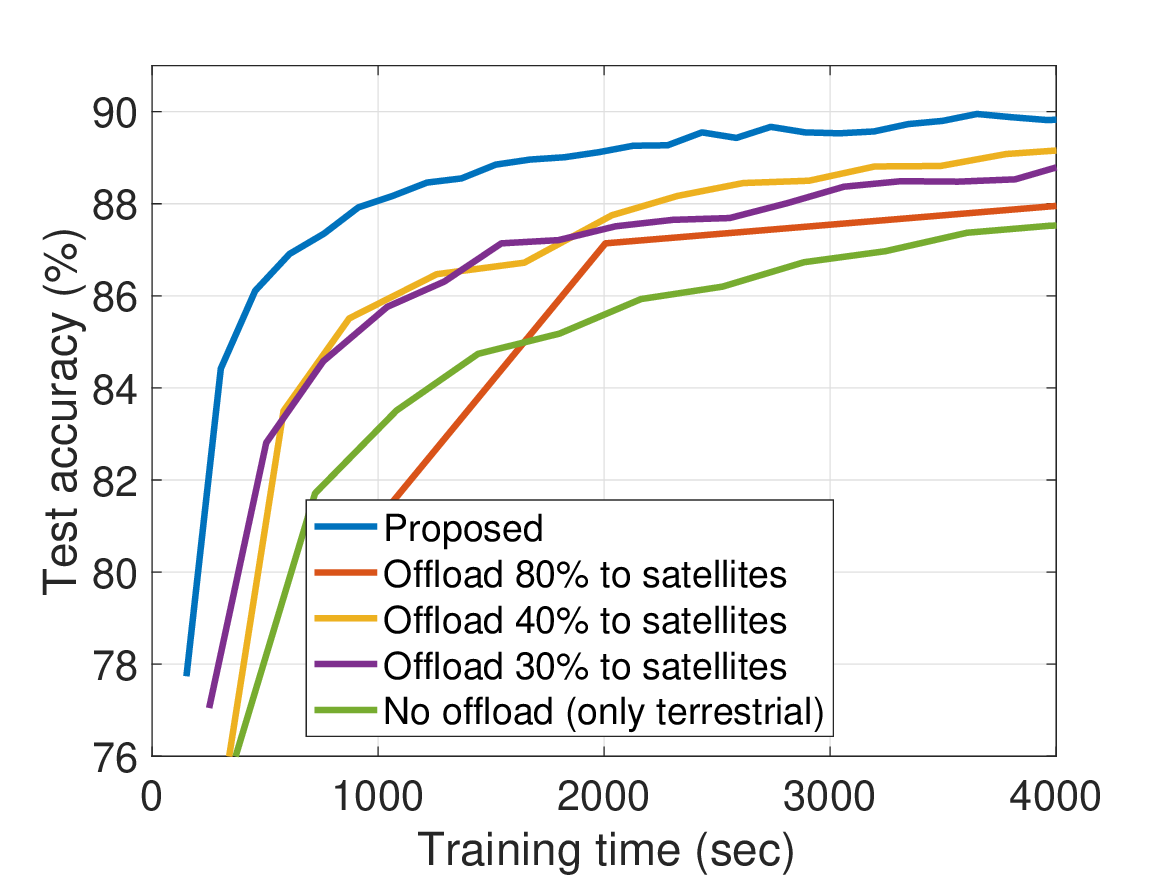}
        \caption{FMNIST}
  \end{subfigure}
     \caption{Experiments with varying coverage times using the   constellation model in Fig. \ref{fig:Modelinggg}.}
    \label{fig:Rebuttal_walker}
      \end{figure} 

\section{Conclusion}\label{sec:conclusion}

In this paper, we proposed a satellite-assisted FL methodology that enables   ground users in remote areas to collaboratively train an ML model, without requiring  terrestrial communication infrastructures.  By strategically taking advantage of satellites as edge computing units, model aggregators, and relays, the proposed methodology can speed up the FL process in ground-to-satellite integrated networks. We theoretically analyzed the convergence behavior of our approach,  and optimized network resources to minimize the latency. Experimental results  confirmed the advantage of the proposed idea compared to baselines, and  provided  insights into our network optimization solutions.   We believe that our solution can provide a new direction to   FL over hybrid terrestrial and  non-terrestrial networks,  where  reducing the training time based on cooperation among ground users and satellites is of paramount importance. One interesting future direction is to extend our  idea to the  ground-air-space three layer network, having devices/sensors on the ground, UAVs or drones  in the sky, and satellites in the space.  

\appendices
\section{Proof of Theorem 1}

Due to the $L$-smoothness, the following holds:
\begin{align}
 \mathbb{E}[F(\mathbf{w}^{r+1})] - \mathbb{E}[F(\mathbf{w}^r)] &\leq   \underbrace{\mathbb{E}[\langle \nabla F(\mathbf{w}^r), \mathbf{w}^{r+1} - \mathbf{w}^r\rangle]}_{\zeta_1^r}\nonumber  \\ &+   \underbrace{\frac{L}{2}\mathbb{E}[\|\mathbf{w}^{r+1}-\mathbf{w}^r\|^2]}_{\zeta_2^r}. \label{eq:proof}
\end{align} 
It can be seen that the term $\mathbf{w}^{r+1} - \mathbf{w}^r$  appears at both terms of (\ref{eq:proof}). We can write
\begin{align}
&\mathbf{w}^{r+1} - \mathbf{w}^r =  \frac{1}{J}\sum_{j=1}^J \bar{\mathbf{w}}_j^{r+1} - \mathbf{w}^r \\&  =- \frac{1}{J}\sum_{j=1}^J\eta_r\tilde{\nabla}F_j(\mathbf{w}^r)  =- \eta_r\tilde{\nabla}F(\mathbf{w}^r)
\end{align}
where $\tilde{\nabla}F(\mathbf{w}^r) = \frac{1}{J}\sum_{j=1}^J\tilde{\nabla}F_j(\mathbf{w}^r)$ and
\begin{align}
&\tilde{\nabla}F_j(\mathbf{w})=
\\ &\frac{\big(\sum\limits_{k\in G_j}\alpha_k|D_k|\big)\tilde{\nabla}\ell_{S,j}(\mathbf{w})  + \sum\limits_{k\in G_j}(1-\alpha_k)|D_k| \tilde{\nabla}\ell_{C,k}(\mathbf{w}) }{\sum\limits_{k\in G_j}|D_k|}.
\end{align}
It can be seen that
\begin{align}
\zeta_1^r & = \mathbb{E}[\langle \nabla F(\mathbf{w}^r), -\eta_r\tilde{\nabla}F(\mathbf{w}^r)\rangle]  \\ & \underset{(a)}{=}-  \mathbb{E}[\langle \nabla F(\mathbf{w}^r), -\eta_r \nabla F(\mathbf{w}^r)\rangle] \\&
 =-\eta_r  \mathbb{E}[\|\nabla F(\mathbf{w}^r)\|^2] \label{eq:A__1}
\end{align}
holds, where $(a)$ is obtained by taking the expectation with respect to the mini-batch. 

Now we focus on  $\zeta_2$. We start by writing
\begin{align}
\zeta_2^r  &= \eta_r^2\frac{L}{2}\mathbb{E}[\|\tilde{\nabla}F(\mathbf{w}^r)\|^2]\nonumber \\
& \underset{(b)}{\leq}  \eta_r^2L \Big(\mathbb{E}[\nabla F(\mathbf{w}^r)\|^2]  +  \mathbb{E}[\|\nabla F(\mathbf{w}^r) -\tilde{\nabla}F(\mathbf{w}^r) \|^2]  \Big)\nonumber\\
&\underset{(c)}{\leq}  \eta_r^2L \Big(\mathbb{E}[\|\nabla F(\mathbf{w}^r)\|^2]\nonumber \\& \ \ \ \  +  \underbrace{\frac{1}{J}\sum_{j=1}^J\mathbb{E}[\|\nabla F_j(\mathbf{w}^r) -\tilde{\nabla}F_j(\mathbf{w}^r) \|^2]}_{\zeta_3^r}  \Big) \label{eq:A2}
\end{align}
where $(b)$ is obtained from $\|a+b\|^2 \leq 2\|a\|^2 + 2\|b\|^2$  and $(c)$ comes from the convexity of $\|\cdot\|^2$. 

To bound $\zeta_3$, we focus on a specific cluster $j$ and analyze $\mathbb{E}[\|\nabla F_j(\mathbf{w}^r) -\tilde{\nabla}F_j(\mathbf{w}^r) \|^2]$. Note that we have 
\begin{align}
\mathbb{E}[\|\nabla &F_j(\mathbf{w}^r) - \tilde{\nabla} F_j(\mathbf{w}^r)\|^2]\leq  \nonumber \\&\frac{\sum_{k\in G_j}\alpha_k|D_k|}{\sum_{k\in G_j}|D_k|}\underbrace{\mathbb{E}[\| \nabla \ell_{S,j}(\mathbf{w}^r)-\tilde{\nabla}\ell_{S,j}(\mathbf{w}^r)\|^2]}_{\text{satellite at cluster} \ j}  \nonumber \\
&  + \frac{1}{\sum_{k\in G_j}|D_k|} \sum_{k\in G_j}\Big((1-\alpha_k)|D_k|  \nonumber \\
& \ \ \ \ \ \ \ \ \ \ \  \times \underbrace{\mathbb{E}[\| \nabla \ell_{C,k}(\mathbf{w}^r) - \tilde{\nabla}\ell_{C,k}(\mathbf{w}^r)\|^2]}_{\text{client} \ k \ \text{at cluster} \ j}\Big), \label{eq:asdffff}
\end{align}
which holds due to the convexity of $\|\cdot\|^2$. 

\textbf{(i) Bounding $\mathbb{E}[\| \nabla \ell_{C,k}(\mathbf{w}^r) - \tilde{\nabla}\ell_{C,k}(\mathbf{w}^r)\|^2]$.} Let   $\tilde{D}_k^{(l, r)} \subset \hat{D}_k^{l}$ be the mini-batch and $\lambda_{C,k}=|\tilde{D}_k^{(l,r)}|\leq (1-\alpha_k)|D_k|$ be the corresponding  mini-batch size at client $k$ at round $r$. We have
\begin{align}
 &\mathbb{E}[\| \nabla \ell_{C,k}(\mathbf{w}^r) - \tilde{\nabla}\ell_{C,k}(\mathbf{w}^r)\|^2]  \nonumber\\&=\mathbb{E}\Big[\Big\|\frac{1}{\lambda_{C,k}^r} \sum_{x\in \tilde{D}_k^{(l,r)}} \nabla \ell(x;\mathbf{w}^r)  -     \frac{1}{|\hat{D}_k^c|} \sum_{x\in \hat{D}_k^c} \nabla \ell(x;\mathbf{w}^r)\Big\|^2\Big] \nonumber \\& \underset{(d)}{=}(1-\frac{\lambda_{C,k}^r}{|\hat{D}_k^c|})\frac{Z_{C,k}}{\lambda_{C,k}^r} \label{eq:sample_mean}
\end{align}
where $(d)$ is obtained by using the variance of sample mean. Specifically, $Z_{C,k}$ in (\ref{eq:sample_mean}) is the variance of the gradients computed with the samples in $\hat{D}_k^c$, which is written as 
\begin{align}
Z_{C,k} &= \frac{1}{|\hat{D}_k^c|-1}\sum_{x\in \hat{D}_k^c}\Big\|   \nabla \ell(x;\mathbf{w}^r)  -  \frac{1}{|\hat{D}_k^c|} \sum_{x'\in \hat{D}_k^c}  \nabla \ell(x';\mathbf{w}^r)     \Big\|^2  \\ &\underset{(e)}{\leq} \frac{2(|\hat{D}_k^c|-1)\rho}{|\hat{D}_k^c|}V_{C,k},  \label{eq:Z_def}
\end{align}
where $(e)$ results from the proof of \cite{chang2023asynchronous} and $V_{C,k}$ is the variance of dataset $\hat{D}_k^c$ defined in (\ref{eq:V_def11}).
By combining the results of (\ref{eq:sample_mean}) and (\ref{eq:Z_def}), we obtain
\begin{align}
&\mathbb{E}[\|\nabla\ell_{C,k}(\mathbf{w}^r)-\tilde{\nabla}\ell_{C,k}(\mathbf{w}^r)\|^2]\\&\leq 2\Big(1-\frac{\lambda_{C,k}^r}{|\hat{D}_k^c|}\Big) \frac{(|\hat{D}_k^c|-1)\rho}{\lambda_{C,k}^r|\hat{D}_k^c|}V_{C,k}. 
\end{align}

\textbf{(ii) Bounding $\mathbb{E}[\|\nabla\ell_{S,j}(\mathbf{w}^r) - \tilde{\nabla}\ell_{S,j}(\mathbf{w}^r)\|^2]$.}   Similarly, for the satellite side,  we obtain
\begin{align}
& \mathbb{E}[\| \nabla \ell_{S,j}(\mathbf{w}^r) - \tilde{\nabla}\ell_{S,j}(\mathbf{w}^r)\|^2] \leq \\ &2\Big(1-\frac{\lambda_{S,j}^r}{|\cup_{k\in G_j}\hat{D}_k^s|}\Big) \frac{(|\cup_{k\in G_j}\hat{D}_k^s|-1)\rho}{\lambda_{S,j}^r|\cup_{k\in G_j}\hat{D}_k^s|}V_{S,j} \label{eq:SSS}
\end{align}
where $V_{S,j}$ is the variance of dataset $\cup_{k\in G_j}\hat{D}_k^s$ defined in (\ref{eq:V_def22}).

Note that $|\hat{D}_k^c| = (1-\alpha_k)|D_k|$ and $|\cup_{k\in G_j}\hat{D}_k^s| = \sum_{k\in G_j}\alpha_k|D_k|$ hold. Now by combining  the results of (\ref{eq:asdffff}), (\ref{eq:sample_mean}), (\ref{eq:SSS}), and then inserting it to (\ref{eq:A2}), we obtain
 
\begin{align}
\zeta_3^r&\leq\frac{2}{  \sum_{j=1}^J \sum_{k\in G_j}|D_k|}\sum_{j=1}^J\bigg(\sum_{k\in G_j}\Big(1-\frac{\lambda_{C,k}^r}{|\hat{D}_k^c|}\Big) \frac{(|\hat{D}_k^c|-1)\rho}{\lambda_{C,k}^r}\nonumber\\& \times V_{C,k}  +\Big(1-\frac{\lambda_{S,j}^r}{|\cup_{k\in G_j}\hat{D}_k^s|}\Big) \frac{(|\cup_{k\in G_j}\hat{D}_k^s|-1)\rho}{\lambda_{S,j}^r}V_{S,j}    \bigg)\label{eq:A3}
\end{align}
By inserting  (\ref{eq:A__1}) and (\ref{eq:A2}) to (\ref{eq:proof}) and choosing $\eta_r\leq \frac{1}{2L}$, we obtain 
\begin{align}
\mathbb{E}[F(\mathbf{w}^{r+1})] - \mathbb{E}[F(\mathbf{w}^r)] \leq -\frac{1}{2}\eta_r\mathbb{E}[\|\nabla F(\mathbf{w}^r)\|^2] + \eta_r^2L\zeta_3^r.
\end{align}
By summing up for all $r=0,1,\dots,R-1$ and and utilizing the result in (\ref{eq:A3}), we obtain the result in Theorem 1, which completes the proof.
\bibliography{Reference_Satellite}
\bibliographystyle{IEEEtran}

\begin{IEEEbiography}[{\includegraphics[width=1in,height=1.25in,clip,keepaspectratio]{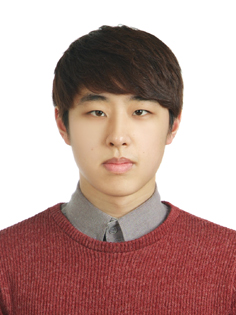}}]{Dong-Jun Han}  (Member, IEEE) received the B.S. degrees in mathematics and electrical engineering, and the M.S. and Ph.D. degrees  in electrical engineering from Korea Advanced Institute of Science and Technology (KAIST), South Korea, in 2016,  2018, and 2022, respectively. He received the Best Ph.D. Dissertation Award from the School of Electrical Engineering at  KAIST   in 2022. He is currently a postdoctoral researcher  in the School of Electrical and Computer Engineering at  Purdue University.  His research interest is at the intersection of communications, networking, and machine learning, specifically in distributed/federated machine learning and network optimization. 
\end{IEEEbiography}

\begin{IEEEbiography}
[{\includegraphics[width=1in,height=1.25in,clip,keepaspectratio]{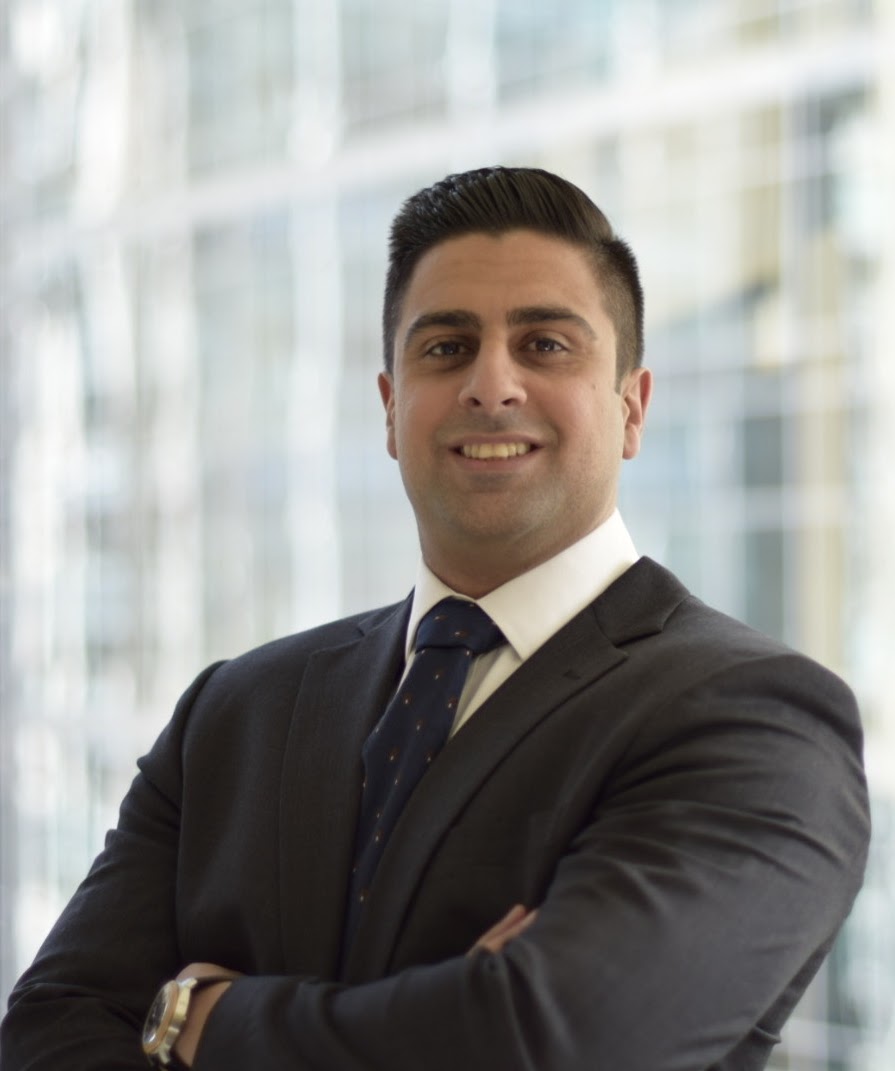}}]{Seyyedali Hosseinalipour} (Member, IEEE)  received the B.S. degree in electrical engineering
from Amirkabir University of Technology, Tehran, Iran, in 2015 with high honor
and top-rank recognition. He then received the M.S. and Ph.D. degrees in electrical engineering from North Carolina State University, NC, USA, in 2017 and
2020, respectively. He was the recipient of the ECE Doctoral Scholar of the Year
Award (2020) and ECE Distinguished Dissertation Award (2021) at North Carolina State University. He was a postdoctoral researcher at Purdue University,
IN, USA from 2020 to 2022. He is currently an assistant professor at the Department of Electrical Engineering at the University at Buffalo (SUNY). He has
served as the TPC Co-Chair of workshops and symposiums related to distributed machine learning and edge computing held in conjunction with IEEE INFOCOM, IEEE GLOBECOM, IEEE ICC, IEEE/CVF CVPR, IEEE MSN, and IEEE VTC. Also, he has served as the guest editor for IEEE Internet of Things
Magazine. His research interests include the analysis of modern wireless networks,
synergies between machine learning methods and fog computing systems, distributed/federated machine learning, and network optimization.
\end{IEEEbiography}

\begin{IEEEbiography}
[{\includegraphics[width=1in,height=1.25in,clip,keepaspectratio]{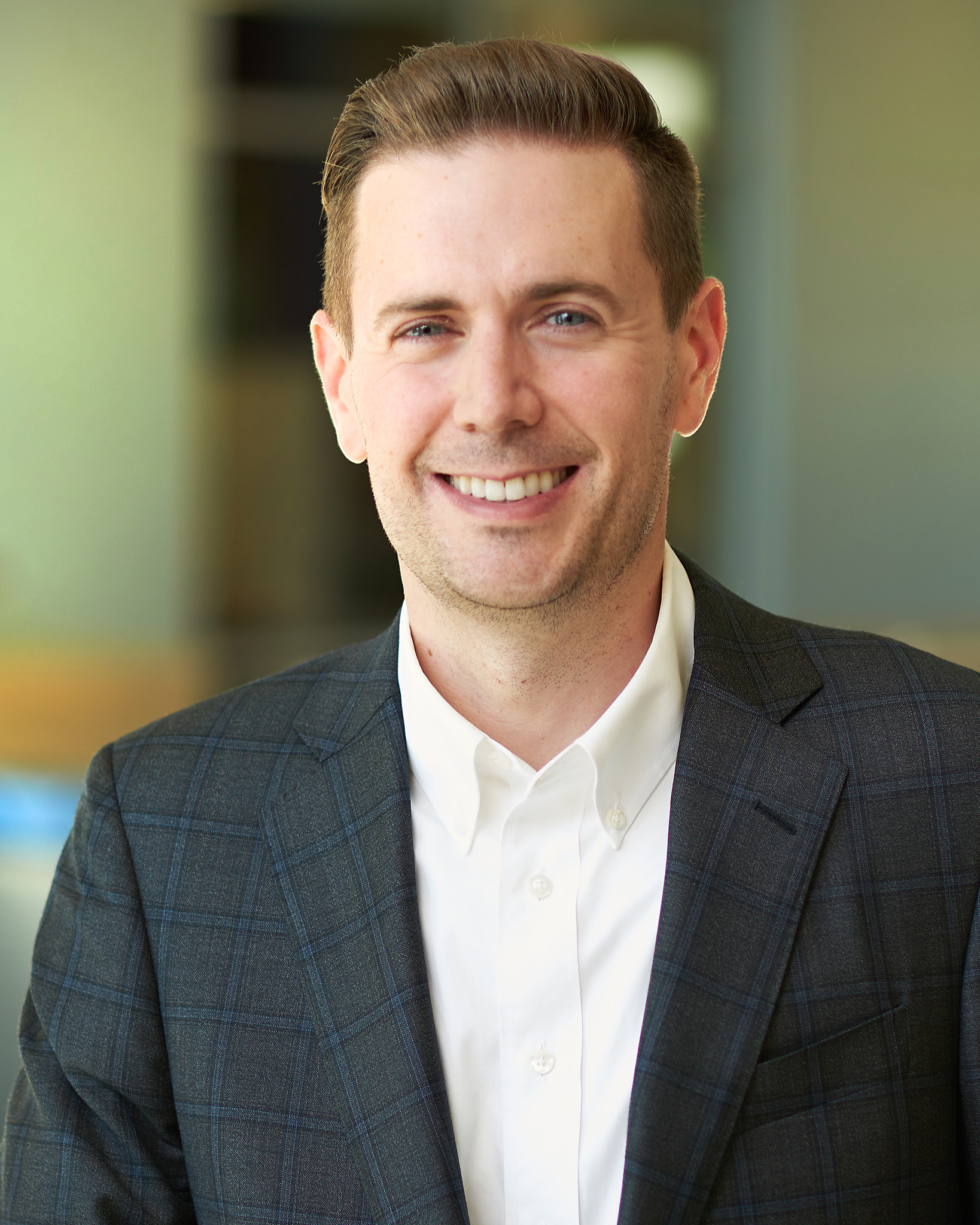}}]{David  J. Love} (Fellow, IEEE) received the B.S. (with highest honors), M.S.E., and Ph.D. degrees in electrical engineering from the University of Texas at Austin in 2000, 2002, and 2004, respectively. Since 2004, he has been with the Elmore Family School of Electrical and Computer Engineering at Purdue University, where he is now the Nick Trbovich Professor of Electrical and Computer Engineering. He served as a Senior Editor for IEEE Signal Processing Magazine, Editor for the IEEE Transactions on Communications, Associate Editor for the IEEE Transactions on Signal Processing, and guest editor for special issues of the IEEE Journal on Selected Areas in Communications and the EURASIP Journal on Wireless Communications and Networking. He was a member of the Executive Committee for the National Spectrum Consortium. He holds 32 issued U.S. patents. His research interests are in the design and analysis of broadband wireless communication systems, beyond-5G wireless systems, multiple-input multiple-output (MIMO) communications, millimeter wave wireless, software defined radios and wireless networks, coding theory, and MIMO array processing. Dr. Love is a Fellow of the American Association for the Advancement of Science (AAAS) and the National Academy of Inventors (NAI).  He was named a Thomson Reuters Highly Cited Researcher (2014 and 2015). Along with his co-authors, he won best paper awards from the IEEE Communications Society (2016 Stephen O. Rice Prize and 2020 Fred W. Ellersick Prize), the IEEE Signal Processing Society (2015 IEEE Signal Processing Society Best Paper Award), and the IEEE Vehicular Technology Society (2010 Jack Neubauer Memorial Award).
\end{IEEEbiography}

\vskip -2\baselineskip plus -1fil

\begin{IEEEbiography}
[{\includegraphics[width=1.0in,height=1.28in,clip,keepaspectratio]{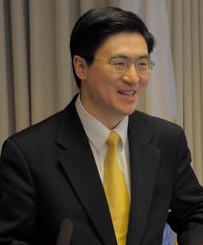}}]{Mung Chiang} (Fellow, IEEE) received the B.S.  degree in
electrical engineering and mathematics and the M.S. and Ph.D. degrees in electrical engineering from Stanford University in 1999, 2000,
and 2003, respectively. He is currently the President of Purdue University and a Roscoe H. George Distinguished Professor of ECE.  
During 2019–2020, he served as the Science and
Technology Adviser to the U.S. Secretary of State
and the chief global technology office in the Department of State to launch Technology Diplomacy. Prior to 2017, he was the
Arthur LeGrand Doty professor of electrical engineering, the inaugural
chair of Princeton Entrepreneurship Council and director of Keller Center
for Engineering Education with Princeton University, Princeton, New Jersey. His research on communication networks received the 2013 Alan T.
Waterman Award, the highest honor to scientists and engineers under the
age of 40 in the U.S. He is a recipient of a Guggenheim Fellowship and the IEEE
Tomiyasu Technical Achievement Award, and he was elected to the National
Academy of Inventors and the Royal Swedish Academy of Engineering Sciences. He founded the Princeton EDGE Lab, in 2009 and co-founded several startup companies and an industry consortium in mobile networks, IoT
and AI. He is a recipient of the ASEE Terman Education Award, and his textbooks
and online courses have reached hundreds of thousands of students.
\end{IEEEbiography}

\vskip -2\baselineskip plus -1fil

\begin{IEEEbiography}
[{\includegraphics[width=1in,height=1.25in,clip,keepaspectratio]{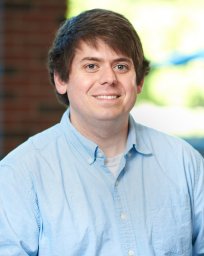}}]{Christopher G. Brinton} (Senior Member, IEEE) is the Elmore Rising Star Assistant Professor of Electrical and Computer Engineering (ECE) at Purdue University. His research interest is at the intersection of networking, communications, and machine learning, specifically in fog/edge network intelligence, distributed machine learning, and data-driven wireless network optimization. Dr. Brinton is a recipient of the NSF CAREER Award, ONR Young Investigator Program (YIP) Award, DARPA Young Faculty Award (YFA), Intel Rising Star Faculty Award, and roughly \$15M in sponsored research projects as a PI or co-PI. He has also been awarded Purdue College of Engineering Faculty Excellence Awards in Early Career Research, Early Career Teaching, and Online Learning. He currently serves as an Associate Editor for IEEE/ACM Transactions on Networking. Prior to joining Purdue, Dr. Brinton was the Associate Director of the EDGE Lab and a Lecturer of Electrical Engineering at Princeton University. He also co-founded Zoomi Inc., a big data startup company that has provided learning optimization to more than one million users worldwide and holds US Patents in machine learning for education. His book The Power of Networks: 6 Principles That Connect our Lives and associated Massive Open Online Courses (MOOCs) have reached over 400,000 students to date. Dr. Brinton received the PhD (with honors) and MS Degrees from Princeton in 2016 and 2013, respectively, both in Electrical Engineering. 
\end{IEEEbiography}
\end{document}